\documentclass[12pt]{article}
\usepackage{graphicx}
\usepackage{color}
\usepackage{amsmath,slashed, amssymb}
\usepackage{fancyhdr}

\def\beq{\begin{eqnarray}}
\def\eeq{\end{eqnarray}}
\def\bea{\begin{eqnarray*}}
\def\eea{\end{eqnarray*}}
\def\bal{\begin{align}}
\def\eal{\end{align}}

\def\centeron#1#2{{\setbox0=\hbox{#1}\setbox1=\hbox{#2}\ifdim
\wd1>\wd0\kern.5\wd1\kern-.5\wd0\fi
\copy0\kern-.5\wd0\kern-.5\wd1\copy1\ifdim\wd0>\wd1
\kern.5\wd0\kern-.5\wd1\fi}}
\def\ltap{\;\centeron{\raise.35ex\hbox{$<$}}{\lower.65ex\hbox{$\sim$}}\;}
\def\gtap{\;\centeron{\raise.35ex\hbox{$>$}}{\lower.65ex\hbox{$\sim$}}\;}

\def\singleandthirdspaced{\baselineskip=\normalbaselineskip\multiply
    \baselineskip by 130\divide\baselineskip by 100}

\newcommand{\newc}{\newcommand}
\newc{\qbar}{{\overline q}}
\newc{\Kahler}{K\"ahler }
\newc{\deltaGS}{\delta_{\rm GS}}

\newcommand{\lp}{\lambda_{p}}
\newcommand{\Fg}{\mathcal{F}_G}
\newcommand{\Fgp}{\mathcal{F}_{G,\,p}}
\newcommand{\Fgpd}{\overline{\mathcal{F}}_{G,\,p}}
\newcommand{\Dgp}{D_{G,\,p}}
\newcommand{\Dg}{D_G}
\newcommand{\pp}{\phi_{p}}
\newcommand{\Fp}{F_{p}}

\begin{document}
\begin{titlepage}
\begin{flushright}
\end{flushright}

\vskip 1.2cm

\begin{center}

{\LARGE\bf (Non-)Decoupled Supersymmetric Field Theories}

\vskip 1.4cm

{\large  Lorenzo Di Pietro,$^{(1)}$ Michael Dine,$^{(2)}$ and Zohar Komargodski$^{(1)}$ }
\\
\vskip 0.4cm

{\it $^{(1)}$Department of Particle Physics and Astrophysics,\\ Weizmann Institute of Science, Rehovot 76100, Israel } \\
{\it $^{(2)}$Santa Cruz Institute for Particle Physics and
\\ Department of Physics,
     Santa Cruz CA 95064  } \\

\vskip 4pt

\vskip 1.5cm

\begin{abstract}

\noindent We study some consequences of coupling supersymmetric theories to (super)gravity. To linear order, the couplings are determined by the energy-momentum supermultiplet. At higher orders, the couplings are determined by contact terms in correlation functions of the energy-momentum supermultiplet. We focus on the couplings of one particular field in the supergravity multiplet, the auxiliary field $M$. We discuss its linear and quadratic (seagull) couplings in various supersymmetric theories. In analogy to the local renormalization group formalism \cite{Osborn:1989, Jack:1990, Osborn:1991gm}, we provide a prescription for how to fix the quadratic couplings. They generally arise at two-loops in perturbation theory. We check our prescription by explicitly computing these couplings in several examples such as mass-deformed $\mathcal{N}=4$ and in the Coulomb phase of some theories. These couplings affect the Lagrangians of rigid supersymmetric theories in curved space. In addition, our analysis leads to a transparent derivation of the phenomenon known as Anomaly Mediation. In contrast to previous approaches, we obtain both the gaugino and scalar masses of Anomaly Mediation by relying just on {\it classical, minimal} supergravity and a manifestly {\it local and supersymmetric} Wilsonian point of view. Our discussion naturally incorporates the connection between Anomaly Mediation and supersymmetric $AdS_4$ Lagrangians. This note can be read without prior familiarity with Anomaly Mediated Supersymmetry Breaking (AMSB).


\end{abstract}

\end{center}

\vskip 1.0 cm

\end{titlepage}
\setcounter{footnote}{0} \setcounter{page}{2}
\setcounter{section}{0} \setcounter{subsection}{0}
\setcounter{subsubsection}{0}

\singleandthirdspaced

\section{Definition of the Problem and Summary}

\begin{figure}
\center
\includegraphics[scale=1.5]{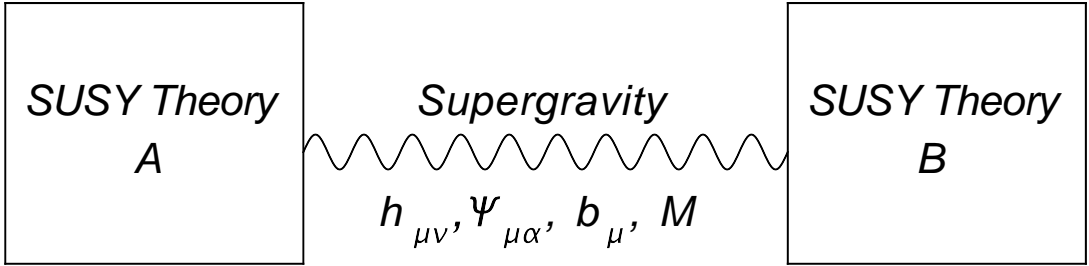}
\caption{The two theories $A$ and $B$ interact through supergravity fields. When we take $M_{Pl}\rightarrow\infty$, the two theories decouple.}
\label{sugramediation}\end{figure}

Consider a supersymmetric
quantum field theory (QFT) with two sectors, $A$ and $B$. {\it Our central assumption is that in the limit of $M_{Pl}\rightarrow\infty$ the theories $A$ and $B$ are exactly decoupled.} 
With this assumption, at energies much below $M_{Pl}$, the theories $A$, $B$ only communicate via supergravity fields and via irrelevant operators induced by Planck-scale physics. The supergravity fields may be embedded in the (old-)minimal supergravity multiplet, which consists of the metric field $g_{\mu\nu}$, the gravitino $\Psi_{\mu\alpha}$, a vector field $b_\mu$, and a complex scalar $M$. (In the minimal supersymmetric Einstein-Hilbert action, the fields $b_\mu$ and $M$ are non-propagating. However, they are still important since they can couple to matter fields and induce various interactions.) See figure~\ref{sugramediation}.

For example, if both theories $A$ and $B$ are asymptotically free, then we can imagine quartic terms such as $\int d^4\theta \frac{c}{M_{Pl}^2}{Q_A}^\dagger Q_A \,{Q_B}^\dagger Q_B$, where $Q_{A,B}$ are some chiral fields of dimension~1 in the theories $A$, $B$ respectively, and $c$ is some order 1 coefficient.  Such local terms can be induced by unknown Planck-scale physics and the parameter $c$ is therefore incalculable.  In addition to such incalculable terms, there may be
genuine calculable low-energy interactions mediated by the (old-)minimal supergravity multiplet~$(g_{\mu\nu},\Psi_{\mu\alpha},b_\mu, \,M)$. 

One can suppress the incalculable effects from Planck-scale physics by assuming that the dimensions of certain operators as measured in the UV SCFTs associated to $A$ and $B$ are high enough. For example, let us imagine that the theory $A$ is the Supersymmetric Standard Model (SSM) (not necessarily its minimal version). Then the theory $B$ can be treated as some ``hidden sector." Let us now suppose that the UV SCFT associated to $B$  has no non-chiral operators of dimension $\leq 2$. (This in particular means that there are no global non-R symmetries in the UV SCFT. This assumption may be somewhat relaxed, but we still make it  for simplicity.)  This assumption about $B$ is sometimes referred to as ``conformal sequestering''\cite{oai:arXiv.org:hep-th/0105137, oai:arXiv.org:hep-th/0111231}. 

With the assumption of conformal sequestering, there is a simple physical question to which the answer must be given by analyzing the exchange of the minimal supergravity fields at energies much below $M_{Pl}$ and, in particular, the answer must be finite. Assuming SUSY is broken in the theory $B$, one can ask what is the SUSY-breaking mass-squared that is mediated to the squarks and sleptons of the SSM. Due to the assumption of conformal sequestering, the irrelevant operators induced by unknown Planck-scale physics connecting the visible quark or lepton superfields with the hidden sector must be of the type 
\beq\label{gm} \int d^4\theta \frac{QQ^\dagger {\cal O}_B}{M_{Pl}^{\Delta_{{\cal O}_B}}}~,
\eeq
where $Q$ is a quark or lepton superfield in the SSM, and  ${\cal O}_B$ is an operator in the UV SCFT of $B$. 
We see that such irrelevant terms are suppressed by $M_{Pl}^{\Delta_{{\cal O}_B}}$ and $\Delta_{{\cal O}_B}>2$. Hence, one immediately concludes that the contribution to the mass squared at order $M_{Pl}^{-2}$ is calculable and is dominated by small momenta (a priori, it could be zero).

\begin{figure}
\center
\includegraphics[scale=1.5]{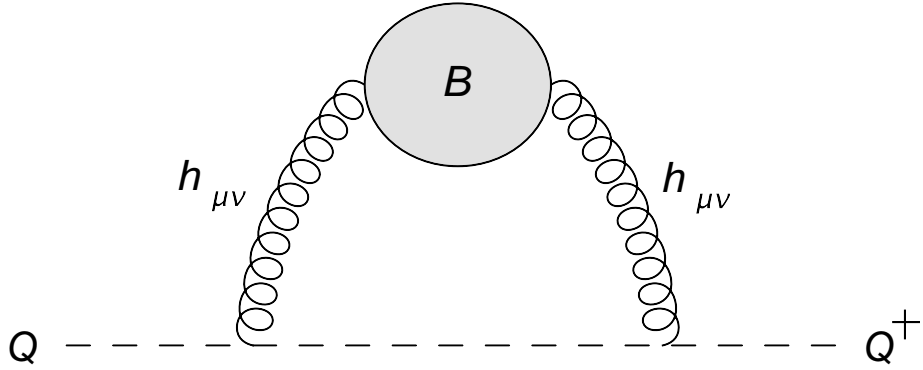}
\caption{Processes contributing to the visible mass squared of sleptons and squarks. Naively, they are  of order $M_{Pl}^{-4}$. }
\label{gravitonloop}\end{figure}

Typical low-energy supergravity interactions coupling the two sectors and potentially inducing a mass squared for the squarks and sleptons involve diagrams such as the one in figure~\ref{gravitonloop}. Since each vertex contributes $M_{Pl}^{-1}$, this diagram can be naively estimated as $M_{Pl}^{-4}$. These diagrams could diverge,  but one must keep in mind that there are no counter-terms of the order $M_{Pl}^{-2}$.   
Similarly, the gaugino masses of the SSM can be induced by UV counter-terms of the type 
\beq\label{gmgaugino} \int d^4x\int d^2\theta \frac{W_\alpha^2 {\Phi}_B}{M_{Pl}^{\Delta_{\Phi_B}}}~,
\eeq
where $\Phi_B$ is some chiral operator in the UV SCFT associated to $B$. If we make the assumption that all the chiral operators in $B$ have dimension $>1$,  we find that contributions to the gaugino mass of order $M_{Pl}^{-1}$ are calculable and dominated by small momenta.

As it turns out, even with the assumption of conformal sequestering, processes that generate a non-holomorphic mass squared of order $M_{Pl}^{-2}$ (and gaugino mass of order $M_{Pl}^{-1}$) do exist.\footnote{Note that in some cases, such as weakly coupled string theory, there is new physics much before the Planck scale. Hence, depending on the details, there might be stringy effects that wash out the calculable terms of order $M_{Pl}^{-1}$ and $M_{Pl}^{-2}$. }  To give an intuitive feeling where such effects might come from, it is helpful to recall some facts about the situation when the theories $A$, $B$  only couple through $U(1)$ (super)gauge interactions.  The Lagrangian is:
\bal
{\cal{L}}_{GGM} =&\,-\frac{1}{4g^2}F^2+ \frac 1{2g^2} D^2 - \frac{i}{g^2}\lambda\sigma^\mu\partial_\mu\bar\lambda \nonumber \\
& -A^{\mu}(j_\mu^{(A)}+j_\mu^{(B)})+D(J^{(A)}+J^{(B)})-\left[\lambda^\alpha(j_\alpha^{(A)}+j_\alpha^{(B)})+c.c.\right]+\cdots~,\label{ggm} 
\end{align}
where $(J^{(A)}, \, j_\alpha^{(A)}, j_\mu^{(A)})$, $(J^{(B)}, j_\alpha^{(B)}, j_\mu^{(B)})$ comprise the global symmetry linear multiplets of the theories $A$ and $B$, respectively. The vector multiplet $(A_\mu,\lambda_\alpha,D)$ couples to these two linear multiplets. The $\cdots$ stand for  interactions between the vector multiplet and matter fields that are quadratic in the vector multiplet (they are fixed by gauge invariance).

In general, the processes that communicate between the sectors $A$ and $B$ are analogous to that of figure~\ref{gravitonloop} with the graviton lines replaced by one of $(A_\mu,\lambda_\alpha, D)$. Imagine again that theory A is the SSM and B is some hidden sector that breaks SUSY spontaneously. Then the mass squared contribution to the sleptons and squarks of $A$ due to gauge interactions is calculable.  The analogs of the diagrams of figure~\ref{gravitonloop}  are of order $g^4$ and lead to the familiar gauge-mediated contributions to the scalar masses. These were studied in~\cite{Meade:2008wd}  (also see references therein).

\begin{figure}
\center
\includegraphics[scale=1.5]{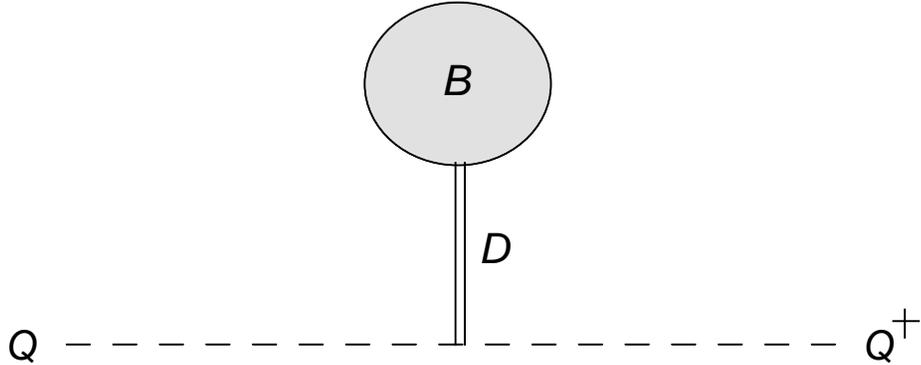}
\caption{When two theories communicate only via gauge interactions, one would have naively thought that the mass squared of the visible sleptons and squarks would be suppressed by $g^4$, but, due to some contact terms that are forced by supersymmetry,  there are also effects of the order $g^2$. }
\label{oneloopggm}\end{figure}

However, there is one notable exception. Suppose the operator $J^{(B)}$ has a nonzero VEV, while the $U(1)$ gauge symmetry is unbroken. Then, from~(\ref{ggm}) we see that from the terms $\frac1{2g^2}D^2+D(J^{(A)}+J^{(B)})$, upon replacing $J^{(B)}$ by its VEV and integrating out the auxiliary field $D$, we find the term $g^2\langle J^{(B)}\rangle J^{(A)}$. Since the bottom component of $J^{(A)}$ is bilinear in squarks and sleptons, this term results in SUSY-breaking scalar masses in the visible sector proportional to $g^2$. This is represented diagrammatically by figure~\ref{oneloopggm}. Note that this is a tree-level effect in the gauge coupling.\footnote{Importantly, the VEV of the bottom component $J$ of a linear multiplet is not a well-defined observable. This is due to an improvement ambiguity that allows to shift the operator $J$ by a constant (the most general improvement takes the form $J \to J + c + (D^\alpha \chi_\alpha + c.c.)$, where $c$ is real and $\chi_\alpha$ is chiral).  When the symmetry is gauged, by coupling the linear multiplet to a dynamical vector multiplet, the VEV becomes a physical observable, namely a Fayet-Iliopoulos (FI) term. However, when the gauge coupling is turned on, there is no well-defined distinction between the operators $J_A$ and $J_B$, since $D$ couples only to the sum of the two, and one may question the validity of the picture of ``mediation of supersymmetry breaking from B to A". Indeed, one can give up this interpretation and take the point of view that the only relevant effect is the coupling of $D$ to an FI term, which results in a non-zero VEV for $D$ which in turn generates the soft masses. This is not always necessary, because typically it is possible to fix the improvement ambiguity before gauging the symmetry, if some additional assumption is satisfied. For instance, the ambiguity is resolved if the theory admits a messenger parity symmetry that acts on the current as $J\to -J$. More relevant for our discussion, we can fix the ambiguity by requiring $\langle J\rangle = 0$ in vacua that preserve supersymmetry and the global symmetry (this is not always possible if there is nontrivial topology in target space, but we ignore this issue here). This is the choice we have made implicit in our discussion above. With this choice, any one-loop contribution to the scalar masses is naturally interpreted as mediated from a nonzero $\langle J_B \rangle$ in the hidden sector. Examples of models realizing this scenario are~\cite{Dimopoulos:1996ig, Argurio:2012qt}.}

Thus, when the theories $A$ and $B$ interact via gauge interactions only, our naive intuition that the scalar masses must be of order $g^4$ is generally incorrect. SUSY forces us to introduce  the contact term $D^2$ in the Lagrangian and the auxiliary field $D$ is forced to couple to an operator in the hidden sector $B$ that may have a nonzero VEV. The latter effect, namely the coupling of $D$ to the bottom component of the linear multiplet, is already present in the limit that the gauge interactions are non-dynamical.

The source of the enhancement of the scalar and gaugino masses in the case that two theories interact only gravitationally is similar. Super-coordinate invariance imposes certain contact term in the coupling of gravity to matter fields and in the kinetic Lagrangian of gravity. These terms communicate supersymmetry breaking from the hidden sector $B$ to the SSM. 

Let us now briefly summarize our findings. The Lagrangian that dictates how the two sectors $A$ and $B$  communicate is as follows:
\begin{align}\label{linearizedgravity}
& {\cal{L}}  = M_{Pl}^2\left(-\frac 12 h_{\mu\nu}E^{\mu\nu} + \epsilon^{\mu\nu\rho\sigma}\bar \Psi _\mu \bar\sigma_\nu\partial_\rho \Psi_\sigma + \frac 13 b_\mu^2- \frac 13 |M|^2\right) \nonumber\\  & + \frac 1 2 \,h^{\mu\nu}\left(T_{\mu\nu}^{(A)}+T_{\mu\nu}^{(B)}\right) + \frac i2 \,\left[\Psi^{\mu\alpha}\left(S_{\mu\alpha}^{(A)}+S_{\mu\alpha}^{(B)}\right)+c.c.\right] \nonumber\\& -\frac 14 \left[M(x^{(A)}+x^{(B)})^\dagger+c.c.\right] - \frac 12 b^\mu\left(j_{\mu}^{FZ; (A)}+j_{\mu}^{FZ; (B)}\right)  +\cdots ~,
\end{align}
We denote the fluctuation of the metric by $h_{\mu\nu}$, and $E_{\mu\nu}$ is the linearized Einstein tensor. The fields $(h_{\mu\nu}, \,\Psi_{\mu\alpha},b_\mu, M)$ comprise the (old-minimal) supergravity multiplet. The operators to which the supergravity fields couple are  $(T_{\mu\nu}, S_{\mu\alpha}, j_\mu^{FZ}, x)$. These form the so-called Ferrara-Zumino multiplet (the vector $j_\mu^{FZ}$ is not generally conserved). The superscripts $(A)$, $(B)$ indicate the sector to which the operators belong. Crucially, in equation~(\ref{linearizedgravity}) we have neglected couplings to matter fields that are quadratic in the supergravity fields and higher. We have also neglected terms that are higher than second order in the kinetic terms. These terms are collectively indicated by the dots.  

In the context that $A$ is the SSM and $B$ is some hidden sector, the assumption that $B$ breaks supersymmetry spontaneously implies the one-point function $\langle T^{(B)}_{\mu\nu}\rangle= -F^2\eta_{\mu\nu}$. To cancel the vacuum energy density we need to take $\langle x^{(B)} \rangle \sim\,F M_{Pl}$ (this leads to an additional, negative,  contribution to the vacuum energy density via~(\ref{linearizedgravity})).\footnote{The ambiguity of the VEV of $J$ in the $U(1)$ analogue, discussed in footnote 2, carries over to this case.  Indeed, while $\langle T^{(B)}_{\mu\nu}\rangle$ is set unambiguously by the dynamics of the hidden sector $B$ (the usual ordering ambiguity is fixed in SUSY theories by requiring that the energy sits on the right hand side of the SUSY algebra), the VEV $\langle x^{(B)}\rangle$ is unobservable in the rigid limit since the operator $X$ is only defined up to $\bar D^2 \Lambda^\dagger$ where $\Lambda^\dagger$ is an anti-chiral superfield (this is the redundancy in the FZ-multiplet described in~\cite{Komargodski:2009pc},\cite{Komargodski:2010rb}; see equations (\ref{improvement}) below). In particular, the constant mode of $x$ is not observable. When we couple to supergravity, only the sum of the VEVs $\langle x_A + x_B\rangle$ is observable.}

Upon integrating out $M$ in~(\ref{linearizedgravity}) we find the contact interaction between the sectors $A$ and $B$: $\sim M_{Pl}^{-2}\,{x^{(B)}}^\dagger x^{(A)}$.
This interaction exists regardless of whether SUSY is broken or not. If SUSY is broken in the sector $B$ we replace ${x^{(B)}}^\dagger$ by its VEV and find the term $\sim M_{Pl}^{-1} F\,x^{(A)}$ in the Lagrangian. All that remains is to calculate the operator $x^{(A)}$ of the visible sector. $x^{(A)}$ sits in the same chiral multiplet as $T^{(A)}=\eta^{\mu\nu}T_{\mu\nu}^{(A)}$, which is nonzero only if the theory is non-conformal. This chiral multiplet is denoted by $X$ and it is a sub-multiplet of the Ferrara-Zumino multiplet. If we take our visible sector to be the supersymmetric $W=\lambda \Phi^3$ theory then $X \sim \beta(\lambda) \Phi^3$. If the visible sector is a gauge theory with some charged matter (and no superpotential for simplicity) then  $X \sim \frac{\beta(g)} g \,W_\alpha^2$. Hence, one can determine in the two cases $x^{(A)} \sim \beta(\lambda)\phi^3$ and $x^{(A)} \sim \frac{\beta(g)} g \lambda_\alpha^2$, respectively (here $\lambda_\alpha$ is the visible gaugino). The above expressions for $X$ hold to all orders in the visible sector. Therefore, plugging these expressions into $M_{Pl}^{-1} F \, x^{(A)}$, we find that the SUSY breaking in $B$ mediates a SUSY-breaking $A$-term $\sim M_{Pl}^{-1} F \, \beta(\lambda)\phi^3 $ in the first case, and a gaugino mass $\sim M_{Pl}^{-1} F \, \frac{\beta(g)} g \lambda_\alpha^2$, in the second case. This is the AMSB contribution \cite{Giudice:1998xp, Randall:1998uk}. 

It is worth pausing to comment on this derivation of the AMSB gaugino masses (or the $A$-term). We see that all we have done is to write the couplings of the supergravity fields to the visible sector and the hidden sector, where we worked in an expansion in $M_{pl}^{-1}$ but to all orders in the field theory couplings (the Ferrara-Zumino multiplet and thus  $X$ exist even in non-Lagrangian theories). The couplings to leading order in $M_{Pl}^{-1}$ are fixed by super-coordinate invariance and, if SUSY is broken, they lead to the anomaly-mediated contributions to gaugino masses and $A$-terms. The only role played by supergravity is the existence of the tree-level term $-\frac 13 M_{Pl}^2|M|^2$ in the Lagrangian. Also note that this derivation involves (unlike the approach of \cite{Bagger:1999rd} and several more recent treatments) only manifestly local and supersymmetric terms in the effective action. 

A derivation of the gaugino soft mass along the lines outlined here is presented for instance in \cite{Weinberg:2000}. An additional remark we want to make here is that the coupling of the supergravity field $M$ to $x$ must be included also when we want to study rigid supersymmetry in curved space. For example, to {\it preserve} supersymmetry in $AdS_4$ we must turn on a constant value of $M$~\cite{Festuccia:2011ws}. The coupling of $M$ to $x$ therefore implies that, for instance, in $W=\lambda\Phi^3$ theory, if we want to preserve SUSY in $AdS_4$, then we need to introduce an $A$-term in the Lagrangian. (There are many other curved spaces in which one needs to turn on a VEV for $M$ \cite{Dumitrescu:2012at}.) 

Let us now consider the SUSY-breaking non-holomorphic sfermion masses. Those clearly do not arise at order $M_{Pl}^{-1}$. In fact, to understand the origin of scalar masses we need to go beyond~(\ref{linearizedgravity}). The crucial effect comes from couplings between supergravity and matter that are quadratic in the supergravity fields.  Fixing the form of the relevant quadratic terms is one of the results of this note. The idea that there exist quadratic couplings between the backgrounds fields and  matter fields is already familiar from the coupling of a conserved current  to a gauge field, where we need to add terms such as $A_\mu^2 |q|^2$ coupling the vector field to charged scalars. This ``seagull term'' coupling is needed in order to cancel a contact term in $\langle j_\mu j_\nu \rangle$ such that this correlation function is conserved both at separated and coincident points. There are similar seagull terms in supergravity. They are completely fixed by demanding super-coordinate invariance, or equivalently, the conservation of the energy-momentum tensor. There is one conceptual difference from the seagull term in gauge theories: $A_\mu^2 |q|^2$ is already necessary at tree-level in order to fix gauge invariance. The seagull term that we exhibit below is only necessary at higher order in perturbation theory. 

Denoting by $\gamma$ the anomalous dimension of the matter fields, we find that we need to add the following seagull term 
\beq\label{scalarmass}{\cal{L}}_{seagull}\supset  \dot\gamma |M|^2 |q|^2~,\qquad \dot\gamma\equiv \frac{d\gamma}{d\log\mu}~. \eeq
{\it This is necessary whenever we couple minimal supergravity to matter, regardless of supersymmetry breaking.}
As emphasized above, the form of this coupling is fixed by some contact terms in the correlators of the energy-momentum tensor multiplet and it is necessary even if supergravity is non-dynamical. It has to be included when analyzing supersymmetric theories in curved space (for a discussion from a different point of view in the case of $AdS_4$, see also \cite{D'Eramo:2012qd, D'Eramo:2013mya}). In flat space, if SUSY is broken in the hidden sector $B$, we need to turn on a VEV for $M$ in order to cancel the cosmological constant. This then leads to the anomaly-mediated SUSY-breaking contribution to non-holomorphic sfermion masses $m^2\sim \dot\gamma\frac{F^2}{M_{Pl}^2}$. Our analysis is again exact to all orders in the coupling constants of the sectors $A$, $B$ and it is to leading nontrivial order in the gravitational coupling. 

Before we discuss how we determined the non-linear seagull term~(\ref{scalarmass}), let us recallseveral issues that one needs to keep in mind:
\begin{itemize}
\item The seagull terms are, by construction, non-universal. The Ferrara-Zumino multiplet is defined only modulo the equations of motion of the rigid theory, but depending on how one chooses to present it using the microscopic fields, the required seagull terms may be different. Indeed, the correlation functions of operators that are equivalent on-shell are identical  at separated points, but they could differ at coincident points and hence, the seagull terms could differ too.  The precise way one realizes the linear couplings~(\ref{linearizedgravity}) in terms of the microscopic fields is ambiguous, and this affects the form of the seagull term. We will see several  examples of such ambiguities in the main body of this note. (The final answers to physical questions are unambiguous.) 
\item The contact term~(\ref{scalarmass}) clearly appears first at two-loops. One  therefore has to compute various correlation functions of the Ferrara-Zumino multiplet at two loops to directly establish its existence. Here we will use a more general approach, that also allows us to determine this contact term to all orders in perturbation theory. Our approach is analogous to \cite{Osborn:1989, Jack:1990, Osborn:1991gm, Nakayama:2013wda, Baume:2014rla}.
\end{itemize}

The derivation of~(\ref{scalarmass}) proceeds along the following lines. First,  consider some  QFT coupled to a background metric field. It is convenient to organize the correlation functions of the energy-momentum tensor in terms of the generating functional $W[g_{\mu\nu}]$. The generating functional is invariant under diffeomorphisms, which leads to the usual energy-momentum conservation Ward identities. If the theory is conformal, then up to local anomalies, $W[e^\Omega g_{\mu\nu}]=W[g_{\mu\nu}]$. If the theory is non-conformal then the generating functional is no longer invariant under $g_{\mu\nu}\rightarrow e^\Omega g_{\mu\nu}$. Suppose our theory is given by perturbing a conformal field theory by some dimension-$4$ operator $\delta S\sim g \int d^4x \mathcal {O}$. In this theory we can consider the $(n+m)$-point functions $\langle T_{\mu_1\nu_1}(y_1)...T_{\mu_m\nu_m}(y_m) \mathcal O(x_1)... \mathcal O(x_n) \rangle$. These are constrained by the requirement that the energy-momentum tensor is conserved. If one insists on that, then one cannot maintain the equation $T_{\mu}^\mu=0$ at separated points. Instead, one finds the operator equation $T_\mu^\mu=\beta(g) \mathcal O$. This is the familiar operatorial trace anomaly. It follows from diffeomorphism invariance. 
It holds for any $g$ and any background metric (up to local anomalies). 

{\it The requirement that $T_\mu^\mu=\beta(g) \mathcal O$ holds up to local anomalies fixes infinitely many seagull terms.} Indeed, let us start from $g=0$. Then, of course, the conformal factor of the metric is decoupled. Once we turn on $g$, the equation $T_\mu^\mu=\beta(g) \mathcal O$ means that we must couple  $h\equiv \eta^{\mu\nu}h_{\mu\nu}$  to the Lagrangian via the contact term $ \beta(g) \int d^4x h \mathcal O$. However, this is not sufficient, and one needs to add further contact terms of higher order in $h$.  After a short computation one finds that one {\it must} add $\dot\beta\int d^4x h^2\mathcal{O}$ (and so forth). Indeed, if we had not added this $\mathcal{O}(h^2)$ term, certain three-point functions of the energy-momentum tensor would have not been consistent with the equation $T_\mu^\mu=\beta(g) \mathcal O$ (and hence, one would violate diffeomorphism invariance). 

The seagull term  $\dot\beta\int d^4x h^2\mathcal{O}$ is reminiscent of the contact term that we claimed in~(\ref{scalarmass}). Indeed, roughly speaking, they are related by supersymmetry.



As a further check of our claims about the coupling of $M$, we consider mass-deformed finite theories, in which the running of the coupling stops above the threshold. In such cases, above the threshold, the conformal symmetry is broken only by classical effects. Hence, the interactions of $M$ are those dictated by the classical supergravity Lagrangian. The linear coupling to the beta-function and the seagull term that appear below the threshold can be obtained by evaluating loop diagrams with supergravity background fields as external legs.
We also discuss the case of gauge theories in the Coulomb phase. They provide another example in which the couplings of $M$ can be determined via a classical analysis, by considering the effective theory below the scale of the VEV.

Let us very briefly compare to some other approaches. Originally \cite{Giudice:1998xp, Randall:1998uk}, the derivation of AMSB proceeded through the ``conformal compensator formalism" for supergravity. Moreover, an explicit loop calculation of the gaugino mass is presented in \cite{Giudice:1998xp} with the choice of a specific regulator, namely the embedding in UV-finite theories. In \cite{Bagger:1999rd} the gaugino soft mass is explained as a supersymmetric completion of certain non-local terms in the effective action, that are fixed by the anomaly in the conformal symmetry and in the superconformal $R$-symmetry.\footnote{The authors of \cite{Bagger:1999rd} also considered subleading corrections to the gaugino mass induced by the K\"ahler potential. We will not discuss those in the present note.} These terms are readily seen to arise in straightforward Feynman diagram computations.  Many works have appeared since then, attempting to clarify the origin of Anomaly Mediation. Just to mention two recent examples (and see references therein): in \cite{Dine:2007me} the AMSB contribution was understood as a contact term required by
supersymmetry.  In particular, in a Higgs phase (see also \cite{Dine:2013nka}), the effective action is local, and the gluino bilinear
is required by the standard action, written in component form.  With massless particles, the contact term was
argued to be analogous to local terms familiar in gauge theories, along the lines of \cite{Bagger:1999rd}.
In either case, this counterterm could be interpreted as a regulator effect that does not decouple. This non-decoupling term breaks flat-space SUSY. If the vacuum is supersymmetric, it is eventually canceled by infrared dynamics.  In \cite{D'Eramo:2012qd, D'Eramo:2013mya} the central idea is that the anomaly mediated contribution (for example, the gaugino mass or the $A$-term) is actually supersymmetric under a deformed SUSY algebra characteristic of $AdS_4$ space.  Several aspects of these derivations have to be contrasted with the main features of our approach, which we now summarize:
\begin{itemize}
\item{We use a minimal set of auxiliary fields, those belonging to the old minimal supergravity multiplet. In the formalism of conformal supergravity, the set of auxiliary fields is extended to include also a chiral superfield $\Phi$, the ``conformal compensator,'' in such a way that the full super-Weyl invariance is recovered. This extended symmetry fixes completely the coupling of $\Phi$. When supersymmetry is broken, $F_{\Phi} \neq 0$ generates the soft terms. This is the essence of the conformal compensator trick, which underlies the original derivation of AMSB \cite{Giudice:1998xp, Randall:1998uk} and many subsequent discussions. In our derivation $\Phi$ is gauge fixed to 0 and the background field $M$ plays a major role instead. The couplings of $M$ are determined by diffeomorphism invariance and supersymmetry. Even though, in principle, setting $\Phi=0$ is merely a gauge choice and hence one does not gain any new information, it is still instructive (and non-trivial) to understand how to extract the physics directly in the minimal supergravity formalism, without using the extended symmetry. }
\item{Our analysis is manifestly supersymmetric throughout. In particular, we do not introduce any non-supersymmetric counterterm. The coupling of the operator $x$ to $M$ and the seagull term (\ref{scalarmass}) are present in the supergravity Lagrangian, and we use supersymmetry to fix their form. Supersymmetry breaking is introduced only upon choosing certain VEVs for the background fields. For instance, supersymmetry breaking in flat space is realized by taking $g_{\mu\nu} = \eta_{\mu\nu}$ and $M \sim M_{Pl}^{-1}F$. Different choices for the background can lead to supersymmetric Lagrangians on curved spaces. As an example, a VEV for $M$ is necessary for supersymmetry in $AdS_4$, and this explains the analogy between AMSB and supersymmetry in $AdS_4$ discussed in \cite{D'Eramo:2012qd, D'Eramo:2013mya}.}
\item{We do not need to include any non-local term in the effective Lagrangian, or rely on some specific UV regulator. Our approach separates the dynamics responsible for breaking the conformal symmetry, which may or may not rely on quantum effects, from the coupling to supergravity, which, for us, is always classical. Said otherwise, it is instructive to imagine that we have already solved the SUSY QFT on $\mathbb{R}^4$ and then imagine coupling to supergravity. In this way, classical breaking of conformal symmetry and quantum breaking of conformal symmetry of the rigid theory on $\mathbb{R}^4$ are treated on the same footing. This leads us to a description of AMSB using only perfectly local and supersymmetric couplings. Equivalently, if we start with the classical supergravity Lagrangian at a certain UV scale $M_1\ll M_{Pl}$, when we integrate out the modes up to $M_2 < M_1$ we generate new local, supersymmertic couplings in the Lagrangian (not present in the classical supergravity Lagrangian), such as $\sim \beta \bar M \lambda_\alpha^2$ and $\sim\dot{\gamma}|M|^2|q|^2$.}
\end{itemize}

Many of the points we make individually are likely known to workers who have investigated this topic.  However, it seems that our presentation leads to a compelling picture which allows to understand the phenomenon in a transparent fashion. 

Before proceeding, we would like to comment on several research directions in which the approach we develop in this note may be a useful starting point. The results presented here are derived following the logic of the local renormalization group formalism \cite{Osborn:1989, Jack:1990, Osborn:1991gm}, which has recently been object of revived interest \cite{Nakayama:2013wda, Baume:2014rla} but whose full range of applications is yet to be explored. Here we make some very preliminary steps in the direction of a supersymmetrized version of this formalism. It would be interesting to fix the full set of equations for the entire supergravity multiplet, and analyze their consequences for the correlators of the energy-momentum tensor multiplet and for the coupling of rigid supersymmetric theories to curved space. In particular, such a formalism seems to be suited to tackle problems in which radiative corrections to the supersymmetric Lagrangian in curved space play an important role. Another application would be to understand the role of anomalies in partition functions on curved space, see e.g. \cite{oai:arXiv.org:1104.4482}. Finally, the seagull term $\dot{\gamma}|M|^2 |q|^2$ that we discuss here and analogous terms involving other supergravity background fields may be relevant in the discussion of dynamics and phases of gauge theories in curved background (e.g. \cite{oai:arXiv.org:1210.5195}).

In the context of possible phenomenological applications, some questions remain to be explored in our formalism. In \cite{D'Eramo:2012qd, D'Eramo:2013mya} the relation between AMSB and the supersymmetry algebra in $AdS_4$ is used to derive a peculiar form for the couplings of the Goldstino to the matter fields, which differs from the naive expectations from ordinary supersymmetry breaking in rigid field theory. Deriving these couplings within our approach may offer a different perspective and lead to a better understanding of this subject. Moreover, in \cite{Bagger:1999rd, D'Eramo:2012qd, D'Eramo:2013mya} several subleading contributions to the soft masses are obtained, based on the supersymmetric completion of some non-local terms in the effective Lagrangian. It would be interesting to try to re-derive these results using a purely local effective action. 

The outline of this note is as follows. In section 2 we describe the classical couplings of the supergravity multiplet to matter fields. We emphasize the role of classical seagull terms and the interplay with the linear couplings.
In section 3 we describe the Ferrara-Zumino multiplet and the linearized coupling to supergravity. Section~4 contains some examples demonstrating the utility of the formalism. This is where we obtain the couplings of the background field $M$ to the gaugino bilinear and to the $A$-terms.  In section~5 we go beyond linearized supergravity and derive~(\ref{scalarmass}). In section~6 we show that the contact terms that we have derived from consistency conditions on the low-energy effective theory indeed arise upon integrating out supersymmetric matter in the concrete examples of finite theories and of theories in the Coulomb phase.  These examples are used to reconcile some of the different approaches to AMSB which we have mentioned.

\section{Matter Fields Coupled to Background Supergravity: Classical Aspects}
Consider a general theory of chiral superfields $\Phi_i$ with superpotential $W(\Phi_i)$ and K\"ahler potential $K(\Phi_i,\Phi_{\bar i}^\dagger)$. Keeping only the bosonic terms, and neglecting couplings suppressed by $M_{Pl}$, which will not matter in our discussion, the minimal coupling to the supergravity multiplet $(h_{\mu\nu}, \Psi_{\mu\alpha}, b_\mu, M)$ leads to the Lagrangian \cite{Festuccia:2011ws} 
\begin{align}\label{classicallag}&\mathcal{L}_{bosonic}=\left(\frac16 R+\frac19 |M|^2-\frac19 b_\mu^2\right)K+K_{i\bar j} 
\left(F^i \bar F^ {\bar j}-\partial_\mu\phi^i\partial^\mu\bar\phi^{\bar j}\right)+F^iW_i+\bar F^{\bar i}
\bar W_{\bar j}\nonumber \\ &-\frac13 M F^i K_i  -\frac13 \bar M  \bar F^{\bar i} K_{\bar i} -\bar M W-
M \bar W -\frac{i}3b^\mu\left(K_i\partial_\mu \phi^i-K_{\bar i}\partial_\mu\bar \phi^{\bar i}\right)~.
\end{align}

Let us start from the massless free field theory, $K=\Phi^\dagger\Phi$ and $W=0$, and let us concentrate on the terms relevant to the couplings to $M$: $\mathcal{L}\supset |F|^2-\frac13 M F \bar \phi - \frac13 \bar M \bar F \phi +\frac 19 |M|^2|\phi|^2$. In conformal field theories we expect $M$ to be exactly decoupled, so let us see how this expectation is borne out. To linear order, $M$ couples to $\bar \phi F$, which is a vanishing operator in the flat space theory. However, despite $\bar \phi F$ being a vanishing operator, it has a nontrivial contact term in its two-point function: $\langle\bar \phi F(x) \phi \bar F(y)\rangle\sim \delta^{(4)}(x-y)\langle\bar\phi(x)\phi(y)\rangle$. Because of this, to ensure the exact decoupling of $M$, we see that we need to add the seagull term $\frac 19 |M|^2|\phi|^2$. Then, $M$ is decoupled. This is a simple demonstration of how seagull terms are fixed by contact terms in certain correlation functions.

More generally, let us take any conformal field theory and couple it to background supergravity.  We can define the generating functional $W[g_{\mu\nu},\Psi_{\mu\alpha} ,b_\mu, M]$ by path integrating over the matter fields.  In conformal field theories the following equation must hold true
\beq\label{conformaldecoupling} \frac{\delta}{\delta M}W[g_{\mu\nu},\Psi_{\mu\alpha} ,b_\mu, M]=0~.\eeq
This is the statement that $M$ is exactly decoupled from the matter theory. We see that to realize this equation in free field theory, we had to add a seagull term $\frac 19 |M|^2|\phi|^2$ because our linear coupling was to an operator that vanishes on-shell, but has nontrivial contact terms. Of course, in free field theory we can  integrate out $F$ in the presence of $M$, which gives $F=\frac13 \bar M \phi$. Plugging this back into the free field Lagrangian, we verify that $M$ disappears altogether. The contact term in $\langle\bar \phi F(x) \phi \bar F(y)\rangle$ is completely equivalent to the fact that, in the presence of $M$, the equation of motion for $F$ is modified and thus we need a seagull term to maintain~(\ref{conformaldecoupling}).\footnote{The decoupling of the conformal factor of the metric works in a very similar fashion.}

Let us consider now a free massive scalar field $K=\Phi^\dagger\Phi$, $W=\frac12 m \Phi^2$. The relevant couplings are \beq\label{massivei}\mathcal{L}\supset |F|^2+F m \phi+ \bar F \bar m  \bar \phi- M \left(\frac12 \bar m\bar\phi^2+\frac13F\bar\phi\right)-\bar M \left(\frac12 m \phi^2+\frac13\bar F\phi\right)+\frac19|M|^2|\phi|^2~.\eeq
Integrating out $F$, we see that now $M$ couples to a non-vanishing operator on-shell, namely $ - \frac16 \bar m \bar \phi^2$. Hence, as expected, the equation~(\ref{conformaldecoupling}) is no longer true. However, there is a useful generalization of this equation. Imagine that $m$ is promoted to a chiral background field. Then it is natural to define the generating functional $W[g_{\mu\nu},\Psi_{\mu\alpha} ,b_\mu, M; m, \psi_m, F_m]$. Now the following equation holds true
\beq\label{massivefreefield} \left( \frac{\delta}{\delta M} + \frac13\bar m \frac{\delta}{\delta \bar F_m} \right) W[g_{\mu\nu},\Psi_{\mu\alpha} ,b_\mu, M; m, \psi_m, F_m]=0~.  \eeq
This equation not only predicts the linear coupling of $M$ to $-\frac16\bar m \bar \phi^2$, but also correctly constrains all the seagull terms. In effect, it means that the partition function $W$ depends on the combination $\frac13 \bar m M - \bar F_m$.\footnote{One can verify the equation~(\ref{massivefreefield}) as follows. Using~(\ref{classicallag}) with $m$ treated as a chiral superfield (that is non-propagating and does not appear in the K\"ahler potential) we can derive the couplings of the background field $F_m$ to supergravity and to matter. In addition to the terms already mentioned in~(\ref{massivei}) one finds only two additional terms, $\frac12F_m\phi^2+\frac12\bar F_m\bar\phi^2$. (There are no terms coupling $F_m$ to $M$.) Then, integrating out the $F$ component of $\Phi$ in~(\ref{massivei}) we  find $\mathcal{L}\supset  -\left(m\phi-\frac13 M\bar\phi
\right)\left(\bar m \bar\phi-\frac13\bar M \phi
\right)+\frac12(F_m- \bar m M) \phi^2+\frac12(\bar F_m - m \bar M) \bar\phi^2+\frac19|M|^2|\phi|^2=-|m|^2|\phi|^2+\left(\frac 12\left (   \bar F_m - \frac13M\bar m\right)\bar\phi^2+c.c.\right)$. We see that the theory only depends on the combination $\frac13 M \bar m - \bar F_m$, confirming~(\ref{massivefreefield}).} 

The linear coupling of $M$ to $- \frac16 \bar m \bar \phi^2$ means that: 
\begin{itemize}
\item If SUSY is broken in flat space (and therefore $M$ has a VEV), we must add a ``holomorphic mass term'' proportional to $\phi^2$ to the Lagrangian. In particular, imagine that SUSY is broken in the hidden sector $B$. In the SSM we often have the so-called $\mu$-term for the two Higgs doublets  $W=\mu H_uH_d$. Then, as a consequence of SUSY breaking in the sector $B$, we need to add a SUSY-breaking $B_\mu$-term to the Lagrangian $\sim \frac{F}{M_{Pl}}\mu H_uH_d$. This is already present at tree level, in spite of the fact that SUSY breaking takes place in some sequestered sector $B$. 
\item If we want to write the free massive scalar theory on curved manifolds such as $AdS_4$ while preserving SUSY, we need to add such a holomorphic mass term. This term preserves SUSY but breaks the R-symmetry. In general, the R-symmetry cannot be preserved on $AdS_4$ unless the underlying theory is conformal (or perhaps has extended SUSY).
\end{itemize}
An important interpretation of  equation~(\ref{massivefreefield}) is that this is a supersymmetric partner of the classical equation of motion $T_\mu^\mu= - 2 |m|^2|\phi|^2+fermions$. The couplings of $M$ to matter are related by supersymmetry to the  couplings of the conformal factor of the metric to matter.

Our final example is the $\phi^3$ theory, $K=\Phi^\dagger \Phi$ and $W=\frac13\lambda\Phi^3$. It is classically conformal. If we couple it to background supergravity we again find the equation~(\ref{conformaldecoupling}). Indeed, despite there being a superpotential, as we can see from~(\ref{classicallag}), the linear coupling of $M$ is to $\bar W+\frac13 K^iF_i$, which vanishes on-shell. The seagull term $\frac19 |M|^2|\phi|^2$ takes care of the modification of the equation of motion of $F$ due to the coupling to background supergravity. Let us now promote the superpotential coupling $\lambda$ to a superfield with components $(\lambda,\psi_\lambda,F_\lambda)$.
Then we can examine the generating functional $W[g_{\mu\nu},\Psi_{\mu\alpha} ,b_\mu, M; \,\lambda, \psi_\lambda, \,F_\lambda]$. One can easily check that at tree level it still obeys~(\ref{conformaldecoupling}), namely, 
\beq\label{conformaldecouplingi}  \frac{\delta}{\delta M}W[g_{\mu\nu},\Psi_{\mu\alpha} ,b_\mu, M; \,\lambda, \,\psi_\lambda, \,F_\lambda]=0~.\eeq
A correction analogous to that in~(\ref{massivefreefield}) is absent due to classical conformal invariance.

In this section we have started from the known classical Lagrangian~(\ref{classicallag}) and extracted the 
equations~(\ref{conformaldecoupling}),(\ref{massivefreefield}),(\ref{conformaldecouplingi}). These equations can be viewed as the physical guiding principles that fix the seagull terms when one couples some matter theory to background supergravity. In the next section we will start developing a more abstract machinery that applies for any SUSY QFT. 

\section{The FZ multiplet and Linearized Supergravity}

In this section we study the linear couplings of arbitrary SUSY QFT to background supergravity. 
We start by reviewing the structure of the Ferrara-Zumino multiplet and its gauging. We follow~\cite{Komargodski:2010rb} (and see references therein). In supersymmetric QFTs, the energy-momentum tensor $T_{\mu \nu}$ and the supercurrent $S_{\mu\alpha}$ are part of a supersymmetric multiplet of operators. This multiplet is not unique, but in all cases it must include some additional operators. A multiplet that always exists unless there are FI terms (or the target space has nontrivial K\"ahler class) is the so-called Ferrrara-Zumino multiplet, which in addition to the supercurrent and the energy-momentum tensor includes a current $j^{FZ}_\mu$ (which is not conserved unless the theory is conformal) and an additional complex scalar operator, $x$. The multiplet can be defined via the superfield equation
\begin{equation}\label{FZconservation}
\bar{D}^{\dot{\alpha}}\mathcal{J}_{\alpha\dot{\alpha}} = D_{\alpha} X~,
\end{equation}
where $\mathcal{J}_{\alpha\dot{\alpha}} = -2\sigma^\mu_{\alpha\dot{\alpha}}\mathcal{J}_\mu$ is a real vector superfield, and $X$ is a chiral superfield $\bar{D}_{\dot{\alpha}}X = 0$. Being an operator equation,~(\ref{FZconservation}) holds in all correlation functions at separated points. 

The expansion of the superfields $\mathcal{J}_\mu$ and $X$ in components gives
\bal
\mathcal{J}_\mu = & \, j_\mu^{FZ} + \left[\theta (S_\mu + \tfrac 13 \sigma_\mu\bar{\sigma}_\nu S^\nu)+ \tfrac i 2 \theta^2 \partial_\mu x^\dagger + c.c.\right]  \nonumber \\
& + \theta\sigma^\nu\bar{\theta}(2 T_{\mu\nu} - \tfrac 23 \eta_{\mu\nu}T + \tfrac 12 \epsilon_{\mu\nu\rho\sigma}\partial^\rho j^{FZ\  \sigma}) + \dots~, \\
X  = & \, x + \tfrac 13 \theta \sigma^\mu S_\mu + \theta^2 (\tfrac 23 T + i \partial \cdot j^{FZ}) + \dots~,
\end{align}
where the dots stand for higher components, and $T=\eta_{\mu\nu}T^{\mu\nu}$. The conservation of the energy-momentum tensor and of the supercurrent follow from (\ref{FZconservation}). One can check that if $j_\mu^{FZ}$ is conserved then the trace of the energy-momentum tensor vanishes, implying superconformal invariance.

Equation (\ref{FZconservation}) does not uniquely determine the multiplet, because given a solution $(\mathcal{J}_\mu, \, X)$ and a chiral operator $\Lambda$, one can find a new solution $(\mathcal{J}'_\mu, \, X')$ by the transformation
\bal\label{improvement}
&\mathcal{J}'_\mu = \mathcal{J}_\mu - i \partial_\mu (\Lambda - \Lambda^\dagger)~, \\ &
X' = X + \tfrac 12 \bar{D}^2 \Lambda^\dagger.
\end{align}
This transformation acts as an improvement transformation on $T_{\mu\nu}$ and $S_{\mu\alpha}$. If a transformation such that $X' = 0$ exists, then the theory is superconformal. 

A special case of~(\ref{improvement}) is a shift of the operator $X$ by a constant. Hence, the expectation value of the bottom component of $X$ is not well defined in the rigid limit. We will momentarily see that it becomes well defined upon gauging the multiplet $\mathcal{J}_\mu$.

The gauging to linear order is done by  coupling the FZ multiplet to the supergravity multiplet $\mathcal{H}_\mu$
\begin{equation}\label{linearcoupling}
\int \, d^4 \theta \, \mathcal{J}^\mu \, \mathcal{H}_\mu~,
\end{equation}
where $\mathcal{H}_\mu$ is a real vector superfield. Using equation (\ref{FZconservation}), we see that this is invariant  under  
\beq\label{gaugesugra}
\mathcal{H}'_{\alpha\dot{\alpha}} = \mathcal{H}_{\alpha\dot{\alpha}} + D_{\alpha}\bar{L}_{\dot{\alpha}} - \bar{D}_{\dot{\alpha}}L_\alpha~,\qquad \bar{D}^2D^\alpha L_\alpha = 0~.
\eeq
 
Equation~(\ref{gaugesugra}) is the linearized supergravity gauge invariance. We can use this gauge invariance to fix a convenient gauge, where the lowest components of $\mathcal{H}_\mu$ vanish, giving the following expansion 
\begin{equation}\label{gravitoncomponents}
\mathcal{H}_\mu =   - \frac 12  \theta\sigma^\nu\bar{\theta}(h_{\mu\nu}-\eta_{\mu\nu} h) + \left[\theta^2 \bar M_\mu + i\bar{\theta}^2\theta(\Psi_\mu+\sigma_\mu\bar{\sigma}_\rho\Psi^\rho) + c.c. \right] - \frac 12 \theta^2\bar{\theta}^2 b_\mu~.
\end{equation}
This is analogous to Wess-Zumino gauge in ordinary gauge theories. 
There is a leftover gauge-invariance acting on~(\ref{gravitoncomponents}) 
\bal
h_{\mu\nu} \rightarrow &\,  h_{\mu\nu} + \partial_\mu\xi_\nu - \partial_\nu\xi_\mu \,, \label{leftoverh} \\
\Psi_{\alpha \mu} \rightarrow & \, \Psi_{\alpha \mu} + \partial_\mu \omega_\alpha \,, \label{leftoverpsi}
\end{align} 
and also $M_\mu$ can be shifted by any conserved vector
\beq\label{leftoverM} M_\mu\rightarrow M_\mu+\chi_\mu~,\qquad \partial^\mu\chi_\mu=0~. \eeq
Equations~(\ref{leftoverh}-\ref{leftoverpsi}) correspond to linearized superdiffeomorphisms and local supersymmetry. The leftover gauge-invariance in $M_\mu$~(\ref{leftoverM}) means that only $M \equiv 2 i\partial^\mu M_\mu$ will appear in Lagrangians.

As always, the gauge-fixed superfield~(\ref{gravitoncomponents}) does not transform well under ordinary supersymmetry, but if we accompany supersymmetry transformations with appropriate gauge transformations  then~(\ref{gravitoncomponents}) becomes a good superfield.

The expansion of the Lagrangian~(\ref{linearcoupling}) is thus 
\beq\label{expansionsugra}
\int \, d^4\theta  \, \mathcal{J}^\mu \, \mathcal{H}_\mu = \frac 12 \,T^{\mu\nu} h_{\mu\nu} + \frac i 2 \, (S^\mu\Psi_\mu + c.c.) - \frac 12\,  j^{FZ;\,\mu}\,b_\mu - \frac 14\, (x \, \bar M + c.c.)~.
\eeq
This defines the linearized coupling of any SUSY theory (that possesses an FZ-multiplet) to the minimal supergravity multiplet.

Let us also discuss the supergravity kinetic term to second order in the supergravity fields. It can be derived from the following superspace Lagrangian
\beq\label{sugrakinetic}
\mathcal{L}_{kinetic}=M_{Pl}^2\int d^4\theta H^\mu E_\mu~,
\eeq
where $E_{\alpha\dot\beta}=\bar D_{\dot \tau}D^2\bar D^{\dot\tau}H_{\alpha\dot\beta}+\bar D_{\dot\tau}D^2\bar D_{\dot\beta}H_{\alpha}^{\dot\tau}+D^\gamma\bar D^2D_\alpha H_{\gamma\dot\beta}-2\partial_{\alpha\dot\beta}\partial^{\gamma\dot\tau}H_{\gamma\dot\tau}$ is a real superfield that is invariant under the gauge transformations~(\ref{gaugesugra}), and it obeys an equation of the type~(\ref{FZconservation}) guaranteeing that the density~(\ref{sugrakinetic}) is gauge invariant.
Expanding~(\ref{sugrakinetic}) in components in our Wess-Zumino gauge we find (up to total derivatives) 
\beq\label{sugrakinetici}
\mathcal{L}_{kinetic}=M_{Pl}^2\left(- \frac 12 h^{\mu\nu}E_{\mu\nu}+ \epsilon^{\mu\nu\rho\kappa}\sigma_{\kappa}^{\dot\alpha\gamma}\bar\Psi_{\mu\dot\alpha}\partial_\nu\Psi_{\rho \gamma} - \frac 13 \, |M|^2 + \frac 13 \, b_\mu^2 \right)~,
\eeq
where $E_{\mu\nu}$ is the linearized Einstein tensor. We see that the fields $M$ and $b_\mu$ are non-propagating.

If SUSY is broken at the scale $\sqrt{F}$, then the rigid theory generates  vacuum energy density $F^2$, which needs to be canceled in order to remain in flat space. The Lagrangians~(\ref{expansionsugra}),(\ref{sugrakinetici}) provide a mechanism for doing so: one declares that $x$ has a VEV (up to a phase)
\beq
\langle x\rangle=\frac{4}{\sqrt 3}FM_{Pl},
\eeq 
which upon integrating out $M$ (which leads to  $ \langle M  \rangle = - \sqrt 3 M_{Pl}^{-1}F$), exactly cancels the energy density generated by the dynamics of the rigid theory. Hence, if SUSY is broken in flat space, $x$ and $M$ have a fixed VEV. 

If one treats gravity as non-dynamical, i.e. one only has~(\ref{expansionsugra}) (and higher order corrections), then coupling the theory  to the $AdS_4$ metric and turning on a constant VEV for $M$ allows to preserve all four supercharges \cite{Festuccia:2011ws}. For the coupling to more general curved spaces, see \cite{Dumitrescu:2012at}.

\section{Examples}

Let us give several simple examples of the FZ multiplet in rigid SUSY theories, and discuss the linearized coupling to supergravity. 

\subsection{Four-Dimensional $\sigma$-Models}
First, following section~2, we re-consider classical four-dimensional $\sigma$-models, i.e.~a collection of chiral superfields $\Phi_i$ with K\"ahler potential $K(\Phi_i,\Phi^\dagger_{\bar i})$ and superpotential $W(\Phi_i)$. One can deduce the Ferrara-Zumino multiplet
\bal\label{nonlinearsigma}
 & \mathcal{J}_{\alpha\dot\alpha}=2g_{i\bar i}D_\alpha \Phi^i\bar D_{\dot\alpha}\bar \Phi^{\bar i}-\frac23[D_\alpha,\bar D_{\dot\alpha}]K~,\nonumber \\ & X =4W-\frac13\bar D^2 K~.
\end{align}
From this we see that $\bar M$ couples at linear order to the combination $4W-\frac13\bar D^2 K$, which, using $\bar D^2K=-4K_{\bar i} \bar F^{\bar i}$, is equivalent to $4(W+\frac13K_{\bar i}\bar F^{\bar i})$. This is precisely what we had in~(\ref{classicallag}). One can infer all the other linearized couplings from~(\ref{nonlinearsigma}) as well. 

Let us now discuss quantum effects, limiting ourselves for simplicity to the theory $K=\Phi^\dagger\Phi$ and $W=\frac \lambda 3  \Phi^3 $. Classically, the theory is conformally invariant, and $M$ decouples. Indeed, it is manifest from~(\ref{nonlinearsigma}) that $X=0$ on-shell.
However, conformal symmetry is broken at the quantum level. Hence, $M$  actually couples to the theory via loop corrections. 

At the scale $\mu$ one can describe the theory by $K=Z(\lambda,\mu)\Phi^\dagger\Phi$ and $W=\frac{\lambda}{3} \Phi^3$, where we used the non-renormalization theorem for the superpotential. We define the physical Yukawa coupling $\lambda_p\equiv\lambda/Z^{3/2}$ and the physically normalized field $\Phi_p\equiv\Phi \sqrt Z$. The usual renormalization group argument implies that $\frac{d\log Z}{d\log\mu}= - \gamma(\lambda_p)$, namely, the derivative of $\log Z$ with respect to the scale only depends on the physical coupling at that scale. With these definitions, $\beta(\lambda_p)\equiv \frac {d\lambda_p}{d\log\mu}= \frac32\lambda_p\gamma$. 

The expression for $X$ in the full theory is easily seen to be $X=\frac{4}{9} \beta(\lambda_{p}) \Phi_p^3$. It is therefore nonzero already at one loop. From this we see that $M$ couples to the theory at the linear order via $ - \frac 1 9 \beta(\lambda_{p}) \bar M \phi_p^3+c.c.$. Upon SUSY breaking in flat space in some hidden sector, this leads to the soft $A$-term $\frac 1 3 A\phi_p^3$, with 
\beq
 A =  \frac F {\sqrt 3 M_{Pl}}\beta(\lambda_p)  
\eeq 
This is the origin of the anomaly mediated $A$-terms. In SUSY Lagrangians on curved manifolds with $\langle M \rangle \neq 0$, the same term gives a coupling proportional to $\phi^3$. 

\subsection{Gauge Theory with Matter}
 
Consider a gauge theory based on gauge group $G$ with some chiral matter in representations $R_i$, and assume for simplicity that  the superpotential vanishes, $W=0$. This theory is  classically conformal; however, it is generally not conformal upon including quantum corrections.
Let us normalize the Lagrangian as 
\beq \mathcal{L}=\frac\tau {32 \pi i} \int d^2\theta \, \mathrm{Tr}(W_\alpha^2+c.c.)+\int d^4\theta\sum_i Q_i^\dagger e^V Q_i\,,
\eeq
 with $\tau=\frac{\theta}{2\pi} + i \frac{4\pi} {g^2}$ and $\mathrm{Tr}(T^a\,T^b) = \frac 12 \delta ^{ab}$. With this normalization, the gauge transformation properties of the field strength superfield $W_\alpha$ are independent of the gauge coupling (and therefore the RG scale). Denoting the wave function renormalization function of the $i$'th superfield by $Z_i$ (and assuming for simplicity no off-diagonal terms) we find 
\beq
X=  \frac 1 3 \left( \frac {b_0} {16\pi^2}  \mathrm{Tr}\,W_\alpha^2 - \frac 1 2 \bar D^2 \sum_i\frac{dZ_i}{d\log\mu}  Q_i^\dagger e^V Q_i \right),
\eeq
 where $b_0$ is the one-loop beta function coefficient ($b_0=3T(G)-\sum_i T(R_i)$ with $T(R_i)$  the Dynkin index of the representation $R_i$).

Note that in the classical theory the axial  current  $ Q_i^\dagger e^V Q_i$ is conserved for every $i$. In the full theory it is anomalous, and after plugging back the anomaly equation
one finally finds 
\beq
X= \frac  1 3 \frac{1}{16\pi^2}\left(b_0 + \sum_i T(R_i)\gamma_i\right) \mathrm{Tr}\, W_\alpha^2	
\eeq
 where $\gamma_i= - \frac{d\log Z_i}{d\log\mu}$. The corrections due the matter wave function renormalization $\sum_i T(R_i)\gamma_i$ are loop suppressed compared to $b_0$ and hence they are negligible unless one has in mind a Banks-Zaks-like fixed point. We will therefore neglect them in the following, although they might be important for some applications.

We are now in position to determine the coupling of the gauge+matter theory to the background supergravity field, $M$. From the bottom component of the superfield $X$ we readily find that one has the supersymmetric coupling $ \frac 1{12} \frac {b_0}{16\pi^2} \bar M \,\mathrm{Tr} \lambda_\alpha^2+c.c.$. The consequences of adding such a term to the supersymmetric $AdS_4$ Lagrangian have been discussed in detail in \cite{Gripaios:2008rg}. Upon SUSY breaking in the hidden sector, by substituting the VEV of $M$ and rescaling the vector superfield $V \to 2 g V$ to have canonically normalized kinetic term, we find the anomaly-mediated gaugino masses
\beq
m_{1/2} =  \frac{b_0 \,g^2}{16\pi^2} \frac F{\sqrt 3 M_{Pl}} .
\eeq

\section{Beyond Linearized Couplings}

In sections~2, 3, 4 we have seen that the linearized couplings of supergravity fields to matter are fixed by the Ferrara-Zumino multiplet of the rigid theory. It is common in the couplings of background fields to matter that one needs to introduce higher order seagull terms in order to maintain various symmetries of the theory. A useful way to keep track of such seagull terms is to define the generating functional $W[\cdot]$ which depends on the background fields, such that derivatives of it lead to consistent correlation functions of the operators to which the background fields couple. 

For example,  coupling a continuous global symmetry associated to a current $j_\mu$ to a background gauge field allows one to consider the generating functional $W[A_\mu]$. One can then {\it define} correlation functions of the operator $j_\mu$ by taking derivatives of $W[A_\mu]$ with respect to $A_\mu$. Current conservation is encoded in the gauge invariance of $W[A_\mu]$.   Note that the correlation functions obtained by taking derivatives of $W[A_\mu]$ with respect to $A_\mu$ are not necessarily identical to what one would have obtained by directly evaluating the correlation functions of $j_\mu$ (realized through the microscopic fields) with, say, Feynman rules. Indeed, $W[A_\mu]$ also receives coincident-points contributions from seagull terms etc., which are necessary in order to have $j_\mu$ appropriately conserved at coincident points. 

Our situation is very similar. As we have seen in section~2, the seagull terms that appear in coupling matter to background gravity are there to make sure that various Ward identities such as~(\ref{conformaldecoupling}),(\ref{massivefreefield})  are respected.  It thus only remains to identify the right generalization of~(\ref{conformaldecoupling}),(\ref{massivefreefield}) and use it to fix the nonlinear coupling of $M$ to matter theories. In this paper we will not attempt to fix all the seagull terms, but just the one involving $|M|^2$.

Let us take a theory with a Ferrara-Zumino multiplet, $\mathcal{J}_{\alpha\dot\alpha}$, satisfying the usual  equation $\bar D^{\dot\alpha}\mathcal{J}_{\alpha\dot\alpha}= D_\alpha X$. 
If $X=0$ on-shell then the theory is conformal and one should impose 
\beq\label{conformalsecfive} \frac{\delta}{\delta M}W[g_{\mu\nu}, \Psi_{\mu\alpha}, b_\mu , M]=0~.\eeq
This means that $M$ is physically decoupled. As we have demonstrated in section~2, one may need to add seagull terms proportional to $|M|^2$ to ensure the exact decoupling. The operatorial meaning of~(\ref{conformalsecfive}) is that $X=0$ also holds at coincident points (in the class of correlation functions that one can obtain from $W[g_{\mu\nu}, \Psi_{\mu\alpha}, b_\mu, M]$).\footnote{Due to anomalies, one might expect that the right hand side of equation~(\ref{conformalsecfive}) is corrected by a {\it local} polynomial in the background fields. (For example, the equation $g_{\mu\nu}\frac{\delta}{\delta g_{\mu\nu}}W[g_{\mu\nu}, \Psi_{\mu\alpha}, b_\mu , M]=0$, which states the decoupling of the conformal factor of the metric in conformal field theories, is clearly corrected by the trace-anomaly polynomial. This is related by supersymmetry to the equation~(\ref{conformalsecfive}).) Whether this is the case or not, does not influence the derivation of the seagull terms (this is because the seagull terms affect correlation functions where some of the points are coincident and some are potentially separated, while the anomaly polynomial is strictly local). Therefore, for our purposes, we may henceforth ignore the anomaly polynomial. In a very closely related setting \cite{Osborn:1989, Jack:1990, Osborn:1991gm, Nakayama:2013wda, Baume:2014rla}, the anomaly polynomial was constrained by dimensional analysis and the Wess-Zumino consistency conditions. For various other applications, it would be very interesting to generalize this to the supersymmetric case.}

Since the theories of interest are not conformal, we seek a generalization of~(\ref{conformalsecfive}). We have already seen an example of how such a generalization might look like in the case of a free massive chiral field~(\ref{massivefreefield}).  In the following, we discuss for concreteness the theory $K=\Phi^\dagger\Phi$ and $W=\frac \lambda 3 \Phi^3$. The case of gauge+matter theories is completely analogous.

As mentioned in section~4,  in this theory $X=\frac 49 \beta(\lambda_p)\Phi_p^3$. (The subscript $p$ stands for `physical', and the physical coupling and field were defined in section~4.)
It will be useful to note that this can be written as $X= \frac 16 Z\gamma \bar D^2 U$, with $U=\Phi^\dagger \Phi$. Hence, the FZ equation becomes $\bar D^{\dot\alpha}\mathcal{J}_{\alpha\dot\alpha}=\frac 1 6 Z\gamma D_\alpha \bar D^2U$. Here $Z$ and $\gamma$ are the wave function renormalization and anomalous dimension functions, respectively.

We can couple our theory to background supergravity: $\int d^4\theta \mathcal J_{\alpha\dot\alpha}\mathcal H^{\alpha\dot\alpha}$. This is invariant on-shell under supergravity transformations $\mathcal{H}_{\alpha\dot\alpha}\rightarrow \mathcal{H}_{\alpha\dot\alpha}+D_\alpha \bar L_{\dot\alpha}-\bar D_{\dot\alpha} L_{\alpha}$, with the constraint $\bar D^2 D^\alpha L_\alpha=0$ (and $D^2\bar D_{\dot\alpha}\bar L^{\dot \alpha}=0$). If we had a super-conformal theory, we would not have to impose any constraint on $L_\alpha$. The constraint that we impose on $L_\alpha$ removes R-gauge transformations, Weyl transformations, and arbitrary shifts of $\mathcal{H}_{\alpha\dot\alpha}\bigr|_{\theta^2}$. (From the latter, only divergence-less shifts remain in non-conformal theories~(\ref{leftoverM}).) 

Let us add sources for the superfield $U=\Phi^\dagger \Phi$. Then, we have at leading order in the background fields $\int d^4\theta\left( \mathcal J_{\alpha\dot\alpha}\mathcal H^{\alpha\dot\alpha}+ G U\right)$, where $G$ is a real superfield that generates correlation functions of $U$.
Now, to linear order in the background fields, the action is invariant under arbitrary (unconstrained) $L_\alpha$ transformations if we accompany the transformation of the background supergravity multiplet $\mathcal{H}_{\alpha\dot\alpha}  \rightarrow \mathcal{H}_{\alpha\dot\alpha}+D_\alpha \bar L_{\dot\alpha}-\bar D_{\dot\alpha} L_{\alpha}$ with an appropriate transformation of $G$. 

The operator $U$ (to which $G$ couples) is a composite operator which undergoes some renormalization. Its renormalization can be understood by viewing $G$  as  the normalization of the kinetic term of the $\Phi^3$ theory in the ultraviolet. The  action in the UV is taken to be $\int d^4\theta G\Phi^\dagger \Phi+\left(\int d^2\theta \frac \lambda 3 \Phi^3+c.c.\right)$. One can determine the structure of the effective action at some scale $\mu$ from symmetries (and the non-renormalization theorem)  to be $\int d^4\theta GZ\left(\frac{\lambda^\dagger\lambda}{G^{3}}\ ;\mu\right)\Phi^\dagger \Phi+\left(\int d^2\theta \frac \lambda 3 \Phi^3+c.c.\right)$. Additionally, there are various irrelevant operators which are unimportant for us. One can use this effective action with arbitrary superfields $G$ and $\lambda$, simply because there are no other terms one can write of dimension $4$. From this, one can define the physical coupling $\lambda_p=\frac{\lambda}{(GZ)^{3/2}}$. The dependence of $Z$ on $\mu$ is restricted to be such that $\frac {d \log Z}{d \log \mu}= - \gamma(\lambda_p)$, namely, the $\mu$-derivative is only a function of the physical coupling at that scale.

From the general form of the effective action, $\int d^4\theta GZ\left(\frac{\lambda^\dagger\lambda}{G^{3}}\ ;\mu\right)\Phi^\dagger \Phi+\left(\int d^2\theta \frac \lambda 3 \Phi^3+c.c.\right)$, one can read off  the operator to which $G$  couples at the scale $\mu$. This is defined to be the renormalized operator $U$ at the scale $\mu$. 

Now we are ready to read off the linearized transformation of $G$ under the general supergravity gauge parameter $L_\alpha$.
Indeed, we need a transformation $G\rightarrow G+\delta G$ such that the transformation of the effective action exactly cancels the piece that comes from the coupling to gravity, $Z\gamma \Phi^\dagger\Phi\left(\bar D ^2 D^\alpha L_\alpha + c.c.\right)$.
We find
\begin{align}&
\mathcal{H}_{\alpha\dot\alpha}  \rightarrow \mathcal{H}_{\alpha\dot\alpha}+D_\alpha \bar L_{\dot\alpha}-\bar D_{\dot\alpha} L_{\alpha}+\cdots \label{extendedinvH}~,\\
& G  \rightarrow  G - \frac1 {48} \gamma\left(\tfrac{\partial \log G  Z}{\partial G}\right)^{-1}(\bar D ^2 D^\alpha L_\alpha + c.c.)+\cdots~. \label{extendedinvG}
\end{align}
In both equations above, the $\cdots$ stand for higher order terms in the background fields (i.e. terms that contain, for instance, at least one $G$ and at least one $L_\alpha$ etc.). The term that we included on the right hand side of the second equation, namely, $ \sim \gamma\left(\tfrac{\partial \log G  Z}{\partial G}\right)^{-1}(\bar D ^2 D^\alpha L_\alpha + c.c.)$ already contains in it higher order couplings (since both $\gamma$ and $Z$ both depend in a complicated way on the superfield $G$), which, strictly speaking, should not be trusted before the higher orders are specified.



Suppose that we have computed the partition function of the theory coupled to these sources, $W[\mathcal{H}, G]$. Then~(\ref{extendedinvH}),(\ref{extendedinvG}) lead to the following differential equation in superspace: 
\beq\label{OsbornSUSY}\int d^4x \int d^4\theta \left[\frac{\delta W}{\delta \mathcal{H_{\alpha\dot\alpha}}}( D_\alpha \bar L_{\dot\alpha} + c.c.) - \frac 1 {48} \frac{\delta W}{\delta  G} \gamma\left(\tfrac{\partial \log  G  Z}{\partial  G}\right)^{-1}(\bar D^2 D^\alpha L_\alpha + c.c.)+\cdots\right]= 0~,
\eeq
where the $\cdots$ stand for higher order terms in the background fields. Those higher-order terms are certainly there, and can be classified systematically. For the problem at hand, the task simplifies as one can check that the corrections to~(\ref{OsbornSUSY}) do not influence the equation involving $\delta W/\delta M$, which is what we are after. Furthermore, as we will momentarily see, the equation one gets for $\delta W/\delta M$ from~(\ref{OsbornSUSY}) is renormalization-group invariant, which is an important consistency check.

\subsection{Derivation of the Seagull Term}

Starting from the UV theory, defined by $\int d^4\theta G\Phi^\dagger \Phi+\int d^2\theta\frac \lambda 3 \Phi^3+c.c.$, we flow to the effective theory at scale $\mu$,  $\int d^4\theta GZ\left(\frac{\lambda^\dagger\lambda}{G^{3}}\ ;\mu\right)\Phi^\dagger \Phi+\left(\int d^2\theta \frac \lambda 3 \Phi^3+c.c.\right)$. Treating $G$ as a superfield, we can expand the effective action in components. In addition, we include the most general couplings of $M$ to the operators of the theory, compatibly with $R$-symmetry and dimensional analysis. This leads to \begin{align}\label{Lsources}
\mathcal{L}(\Fgp,\,\Dgp,\,M)& =   (\Fgp \,\bar \Fp\pp +c.c.) +  \Dgp |\pp|^2+|\Fp|^2 +\left(\lp\Fp\pp^2+c.c.\right)\nonumber \\ & + ( c_1 \,\bar M\, \bar \Fp \pp + c.c.) + (c_2 \, M \,\Fgp + c.c.)|\pp|^2 + c_3 \, |M|^2\,|\pp|^2~,
\end{align} 
where the physical couplings $\Fgp$ and $\Dgp$ can be written explicitly in terms of the couplings of the UV theory (i.e. $G, \, \Fg, \, \Dg$ and $\lambda$) using the general form of the effective action. $c_{1,2,3}$ are unknown coefficients  to be determined using~(\ref{OsbornSUSY}). 

Note that we have chosen to write the linear coupling of  $M$ as $\bar M \,\bar\Fp \pp$, rather than $\bar M \,\pp^3$. The two are equivalent after integrating out $\Fp$, and the difference only amounts to a redefinition of the second order terms $c_2$ and $c_3$. This is the inherent ambiguity present in seagull terms that we discussed in the introduction.

Expanding (\ref{OsbornSUSY}) in components and taking the terms which are proportional to $L_\alpha|_{\theta^2\bar\theta}$ (this is the gauge parameter that allows to shift $M$), we find after some algebra
\begin{align}\label{scomponent}
& \mathcal{D}(x) W  = 0~, \nonumber\\
& \mathcal{D}(x)  \equiv \frac{\delta }{\delta \bar M(x)} - \frac 16 \gamma \frac{\delta}{\delta \Fgp(x)} - \left( \frac 16 \gamma - \frac 12  |\lp|^2 \gamma'\right) \Fgpd (x) \frac{\delta }{\delta \Dgp(x)}~. 
\end{align}
Note that even though we originally derived the expression for the differential operator $\mathcal{D}$ in terms of the UV couplings $(\lambda, \, G,\Fg,\Dg)$, it depends on these sources only through the physical couplings $(\lp,\Fgp,\Dgp)$.  This reflects the RG invariance of our equation~(\ref{scomponent}).

By applying $\mathcal{D}(x)\bar {\mathcal{D}}(y)$ to the effective action~(\ref{Lsources}) we have 
\beq
0 = \mathcal{D}(x)\bar{\mathcal{D}}(y)W|_{\{s,\Fgp,\Dgp\} = 0} = |c_1 - \tfrac 16  \gamma|^2 \langle (\bar \Fp\pp)(x) (\Fp\bar\pp)(y) \rangle+~{\rm coincident-points},~
\eeq
where `coincident-points' stands for terms that only have support for $x=y$. Since the correlation function above at separated points is nonzero, the results can only vanish if $c_1 =  \tfrac 1 6 \gamma$. This relation gives the expected linearized coupling, $\sim \bar M \gamma \bar \Fp \pp$, which is equivalent to leading order in $M$ to $\bar M \gamma \lp \pp^3 \sim \beta(\lp)\bar M \pp^3$. This coincides with the results of section~4.

Substituting $c_1= \frac 1 6 \gamma$ in the Lagrangian (\ref{Lsources}), we obtain
\begin{align}
&0 = \mathcal{D}(x)\bar{\mathcal{D}}(y)W = B(x)\,\bar B(y) \langle|\pp|^2(x) |\pp|^2(y)  \rangle_{sources}+~{\rm coincident-points}~, \label{secondderivative}\\
&B(x) \equiv \left(c_2 - \frac 1 6  \gamma + \frac 1 2  |\lp|^2 \gamma^{(')}\right)\Fgpd(x) + \left(c_3 - \frac 1 6  \bar c_2\gamma\right)\bar M (x)~,
\end{align}
where $\langle \dots \rangle_{sources}$ is the correlator in the theory with background sources. The term `coincident-points' again stands for terms that only have support when $x=y$. We can imagine expanding the right hand side of~(\ref{secondderivative}) in the number of sources. The leading pieces comes from taking $\langle|\pp|^2(x) |\pp|^2(y)  \rangle_{sources}=\langle|\pp|^2(x) |\pp|^2(y)  \rangle_{sources=0}$. Then the two-point function is manifestly nonzero and we thus find that to satisfy the equation we have to take $B=0$, and therefore
\begin{align}
c_2 = &   \frac 1 6 \gamma - \frac 1 2  |\lp|^2 \gamma^{(')} \\
c_3 = & \frac 1 6 \gamma\, \bar c_2 =\frac 1 {36}\gamma^2 - \frac 1 {12} |\lp|^2\gamma\,\gamma^{(')}= \frac 1 {36} (\gamma^2 - \dot{\gamma}).
\end{align}
Alternatively, we can act with two additional operators $\mathcal{D}$ and set all the sources to zero to get the same result. This means that the seagull terms $c_{2,3}$ are set by requiring consistency of four-point functions of $x$. 
In fact, were we to carefully follow the `coincident-points' contribution, we could have derived the same results by just using a three-point function. 

To summarize, the linear in $M$ contact term $\sim \beta(\lp) M \pp^3$ follows from consistency conditions on two-point function and the quadratic one in $M$ follows from consistency conditions on three-point functions. While the source $G$ plays an important intermediate role in the derivation, we find that there are some contact terms of $M$ even when $G$ is a constant. Our final Lagrangian is
\beq
\mathcal{L} =  \frac 1 6 \gamma( \bar M\, \bar \Fp \pp + c.c.) + \frac 1 {36} (\gamma^2 - \dot{\gamma}) |M|^2 |\pp|^2~.
\eeq
Integrating out $\Fp$, this is equivalent to 
\beq\label{seagull}
\mathcal{L}  = - \frac 1 9 \beta(\lp)( \bar M \pp^3 + c.c.) - \frac 1 {36} \dot{\gamma} |M|^2 |\pp|^2~.
\eeq
In particular, upon SUSY breaking in some hidden sector (generating the vacuum energy density $F^2$ and a VEV for $M$), the scalar $\phi$ acquires a physical non-holomorphic mass-squared 
\beq
m^2_{s}=  \frac 1 4 \left (\frac{F}{\sqrt {3}M_{Pl}}\right )^2\dot\gamma.
\eeq

\section{A Derivation from Finite Theories and Coulomb Phase}
\label{finitetheories}

In the previous sections we have seen that the couplings of the supergravity background field $M$, at linear level and beyond, are determined by the running of the parameters of the theory. In this section we will consider cases in which conformal symmetry is broken at the classical level by the introduction of some dimensionful parameter (mass term or VEV), with the property that the running of the dimensionless coupling stops at the massive threshold (above or below the threshold, respectively). Therefore, in the appropriate range of scales, the couplings of $M$ are determined just by classical considerations, as in the examples discussed in section 2. Once the couplings of $M$ are determined at some range of scales, we can evolve to other scales. In this way we will show that the expected couplings of $M$ are generated precisely with the coefficients found in the previous sections.  This is therefore an important consistency check of our proposal.

As a first example, let us consider finite $\mathcal{N}=2$ and $\mathcal{N}=4$ supersymmetric gauge theories, deformed by mass terms to $\mathcal{N}=1$ theories. This case was already considered in the original derivation of the anomaly-mediated gaugino mass in \cite{Giudice:1998xp}, but we find it useful to reproduce it here, both to stress the point of view of the coupling to the background field $M$ (which exists independently of SUSY breaking), and to show that it can be extended beyond linear order to derive the seagull term. For example, consider first the $\mathcal{N}=4$ theory and add mass terms, $\frac12 m \Phi^a \Phi^a$,  for all the adjoint fields.   Because the theory is finite, at energies above the typical mass scale, the operator $X$ of the FZ multiplet is given by its classical expression 
\beq
X = 4 \left ( W - \frac 13 \Phi^a {\partial W \over \partial \Phi^a}  \right )={2 \over 3} m  \Phi^a \Phi^a~.
\eeq

The infrared of this theory is pure $\mathcal{N}=1$ super Yang-Mills theory. Thus, at scales below $m $, one finds $X \sim \beta(g)\mathrm{Tr} W_\alpha^2$,
where $\beta(g)$ is the usual beta function. We will only consider the leading one-loop order in $g$ in order to avoid having to discuss the anomaly puzzle  (\cite{ArkaniHamed:1997mj, Dine:2011gd, Yonekura:2012uk}, see also \cite{Novikov:1983uc, Shifman:1986zi} for some background). Similar remarks apply to the finite $\mathcal{N}=2$ theories.  

The lack of wave function renormalization at scales above $m$ implies that the equation~(\ref{massivefreefield}) holds exactly.  This follows from the fact that the operator $X$ receives no quantum corrections. This means that above the scale $m$ the dependence on $M$ is particularly simple: upon promoting $m$ to a superfield,  the physical observables depend on $F_m$ and $M$  only through $\bar F_{m} - \frac13\bar m  M$. Since it is easy to determine the effective action below the scale $m$ as a function of $F_m$, we can therefore infer the couplings of $M$ rather easily. 

Indeed, the dependence on the superfield $m$ of the low-energy effective action follows from the usual non-renormalization theorems, and at the scale $\mu$ the dependence on $m$ is given by $\int d^2\theta \frac 1 4 \frac{b_0}{16 \pi^2}\log(m/\mu) \mathrm{Tr} W_\alpha^2$, where $b_0=3N_c$ for $SU(N)$ gauge group. Corrections that include covariant derivatives acting on $m$ are irrelevant at low energies. From here one can read off the coupling of $M$ to be $\frac 1 {12} \frac{b_0}{16\pi^2} \bar M \mathrm{Tr} \lambda_\alpha^2$, which gives the expected gaugino mass upon SUSY breaking in some hidden sector. 

As an aside, note that in the presence of the background field $M$ the spectrum of the heavy fields is identical to what it would be in the so-called Minimal Gauge Mediation. Indeed, effectively, one can describe the UV spectrum of the heavy fields $\phi^a$ by a Minimal Gauge Mediation spurion $S=m(1 - \frac 13  \theta^2 \bar M )$. In such cases one expects the infrared result to be proportional to the number of messengers we have integrated out. This is given by the number of massive adjoint chiral multiplet $N_a$, each weighted with the Casimir $N_c$ of the representation, giving a total of $N_a N_c$ messengers. If all the adjoint chiral fields of the UV theory get a mass, then $N_a = 3$, consistent with what we have found.   Indeed, all of the results we have described here, such as the form of $X$ at low energies and the coupling of $\lambda \lambda$ to $M$ are readily obtained from simple Feynman diagram computations.  Because the theories are finite, one can proceed in a pedestrian fashion, ignoring issues of regulators and counterterms. 

By slightly complicating things, we can also find the seagull term containing $|M|^2$. Let us suppose we give a mass $m$  to $1\leq N_a < 3$ of the adjoints. In this case the effective action in the infrared is more complicated, as the remaining $3 - N_a$ adjoints have nontrivial dynamics. (We can imagine giving them a much smaller mass, and in the meanwhile study the effective action in the intermediate regime where they can be viewed as massless.) Since we are dealing with UV-finite theories, above the scale $m$ the dependence on $M$ is again fixed by replacing $\bar F_m \rightarrow \bar F_m - \frac13\bar m  M $. It thus only remains to fix the dependence on the superfield $m$ of the effective action below the scale $m$.

Using the usual non-renormalization theorems one finds for the effective action at some scale $\mu < m$ 
\begin{align}S_{eff}& =\int d^4\theta Z	\left(\tau,  m^\dagger m /\mu^2\right)\sum_{a=1}^{3-N_a}\Phi^{\dagger a}e^V\Phi^a\nonumber \\ & +\int d^2\theta \left({\tau \over 32 \pi i} - \frac 1 4 \frac{N_aN_c}{16\pi^2}\log \left(  m / \mu \right)\right)\mathrm{Tr} W_\alpha^2+c.c.~.\label{finiteeffective} \end{align}
Here $\tau$ is the exactly marginal parameter of $\mathcal{N}=4$ theory. (This determines the gauge coupling at and above the scale $m$.) The fact that $Z$ depends only on $m^\dagger m$ is obvious from the symmetries of the theory, and the rest is fixed by dimensional analysis and holomorphy. It also follows from dimensional analysis that the above effective action is the leading result when $m$ is an arbitrary superfield.

From~(\ref{finiteeffective}), we see that the linear contact term for the gaugino mass is now proportional to $3N_c - (3 - N_a) N_c = N_a N_c$, as expected on general grounds.\footnote{This is also proportional to the number of Minimal Gauge Mediation messengers we integrated out.} 

We can now turn to the seagull term containing $|M|^2$. We can read off the result directly from the effective action~(\ref{finiteeffective}). After a short calculation in which one extracts the coefficient of $|\phi|^2|M|^2$ after integrating out $F_\phi$, one finds that it is  proportional to $\dot\gamma$ with the coefficient we found in the previous section.\footnote{The calculation can again be phrased in the language of Minimal Gauge Mediation, with one difference: there are Yukawa couplings linking the light and heavy scalars. Phenomenologically, such terms correspond to ``Yukawa Mediation.''}  This therefore confirms the presence of the seagull term~\eqref{scalarmass} in the coupling to non-dynamical supergravity.

As a second example, we consider some theory (such as pure $SU(2)$ $\mathcal{N}=2$ theory) in the Coulomb phase, where the running of the gauge coupling stops at the scale of the VEV. The theory below the scale of the VEV is therefore classical and its coupling to supergravity is the usual one. The low-energy theory contains a neutral chiral field $u$ and a massless gauge field.  
In order to see how this works, we start with the effective action below the scale $u$ of the VEV, in the absence of the gravitational field, 
\beq
{\cal L} = \int d^2 \theta \left({1 \over 4 g^2} + {b_0 \over 32 \pi^2} \log\left(u/ \Lambda \right)\right) W_\alpha^2,
\eeq
where $\Lambda$ is the cutoff. (We dropped the instanton corrections, which are not important for our purpose since we will mostly discuss $u\gg \Lambda$.)

This nonlinear theory can be coupled to supergravity via the usual classical prescription. In particular, the couplings relevant to us take the form
\beq
\mathcal{L} = |F_u|^2  - \frac 13( F_u M \bar u + c.c. ) - {b_0\over 32\pi^2}{F_u \over u } \lambda_\alpha^2.
\eeq
Upon integrating $F_u$ out, the coupling $\sim {\beta(g) \over g}\bar M \lambda_\alpha^2$ is generated. (We assumed that the K\"ahler corrections are small, which is correct at large $u$.)
We see that a term very similar to our contact term is generated in this theory at low energies via classical supergravity (due to the nontrivial gauge kinetic function). Since its coefficient is independent of $u$, it is tempting to conclude that it reflects a similar (but not visible in classical supergravity) coupling of $M$ to all the gauginos at scales above $u$. The coefficient of this contact terms supports this interpretation.  

Also the seagull term can be derived in this fashion, by taking into account the correction to the K\"ahler potential for the $u$ field. This comes from the wave-function renormalization at the scale $\mu = |u|$. (We again neglect instanton corrections.) Indeed, plugging in equation~(\ref{classicallag}) a non-trivial K\"ahler function of the form $K(u,\,\bar u) = Z(|u|) |u|^2 $ one readily finds the Lagrangian
\beq
\mathcal{L} = Z\left(\left(1-  \frac \gamma 2 \right)^2 -  \frac {\dot{\gamma}} 4  \right)|F_u|^2  - Z\left(1 - \frac \gamma 2\right) \frac 13( F_u M \bar u + c.c. ) + \frac 1 9 Z |M|^2 |u|^2~.
\eeq
When $F_u$ is integrated out, we are left with the seagull term $\sim \dot{\gamma} |M|^2|u|^2$, with the same coefficient as in (\ref{seagull}). This is again present already in classical supergravity, but one is tempted to conclude that it descends from a corresponding quantum contact term involving $|M|^2$ and the adjoint scalar in the UV.

Note that this analysis is very similar to the one of~\cite{Dine:2007me}. The difference is that here we are working in the Coulomb phase rather than the Higgs phase, so the effective action we are using is manifestly well defined.  Ref.  \cite{Dine:2013nka} in fact considered Coulomb phase examples, as well as non-abelian extensions of the Higgs phase examples of ~\cite{Dine:2007me}.

The various approaches to anomaly mediation can all be understood in this framework.  First, it is again worth stressing that the gaugino mass is completely local and supersymmetric.  At a microscopic level, $M$ couples to $x = {2 \over 3}  m \phi^a \phi^a$ (in the $N=4$ case); this is completely as expected from the standard supergravity analysis.  At long distances, $x$ can be replaced by $x = {1 \over 3}{\beta(g) \over g} \lambda \lambda$.  So the superpotential, in the standard supergravity formulas, should be modified, including the gaugino bilinear in $x$.  Indeed, this was the observation of \cite{Dine:2013nka}, where it was noted that in theories in which a superpotential is generated non-perturbatively through gaugino condensation, it is the full superpotential which should appear in the supergravity Lagrangian.

 While the gaugino mass term is completely local and supersymmetric, it is not surprising that many Green's functions in the low energy theory become singular at low momenta (for example, involving external gravitons and gauge bosons).  As in \cite{Bagger:1999rd}, some of these are related by supersymmetric Ward identities to non-singular contact terms.  This is also consistent with the viewpoint in \cite{Dine:2007me}.  

In the deformed finite theories, the conformal compensator approach, not surprisingly, maps on to the treatment of the mass, $m$, as a spurion.  This is closely related to the coupling $G$ in section 5. 

\section*{Acknowledgments}

We are grateful to O. Aharony, R. Argurio, M. Bertolini, P. Draper, G. Festuccia, M. Luty, R. Rattazzi, A. Schwimmer, N. Seiberg, R. Sundrum, and especially J. Thaler for useful discussions. The work of M.D. was supported in part by the U.S. Department of Energy.  M. Dine is grateful to the Weizmann Institute of Science for hospitality during the initial stages of this project. L.D.P. and Z.K. are supported by the ERC STG grant number 335182, by the Israel Science Foundation under grant number 884/11. L.D.P. and Z.K. would also like to thank the United States-Israel Binational Science Foundation (BSF) for support under grant number 2010/629. In addition, the research of L.D.P. and Z.K. is supported by the I-CORE Program of the Planning and Budgeting Committee and by the Israel Science Foundation under grant number 1937/12. Finally, L.D.P. and Z.K. are grateful to KITP for hospitality at the final stages of this project. Any opinions, findings, and conclusions or recommendations expressed in this material are those of the authors and do not necessarily reflect the views of the funding agencies.
~

\newpage

\bibliographystyle{JHEP}
\bibliography{Biblio}

\providecommand{\href}[2]{#2}\begingroup\raggedright\begin{thebibliography}{10}

\bibitem{Osborn:1989}
H.~Osborn, {\it {Derivation of a Four-dimensional $c$ Theorem}},  {\em
  Phys.Lett.} {\bf B222} (1989) 97.

\bibitem{Jack:1990}
I.~Jack and H.~Osborn, {\it {Analogs for the $c$ Theorem for Four-dimensional
  Renormalizable Field Theories}},  {\em Nucl.Phys.} {\bf B343} (1990)
  647--688.

\bibitem{Osborn:1991gm}
H.~Osborn, {\it {Weyl consistency conditions and a local renormalization group
  equation for general renormalizable field theories}},  {\em Nucl.Phys.} {\bf
  B363} (1991) 486--526.

\bibitem{oai:arXiv.org:hep-th/0105137}
M.~A. Luty and R.~Sundrum, {\it {Supersymmetry breaking and composite extra
  dimensions}},  {\em Phys.Rev.} {\bf D65} (2002) 066004,
  [\href{http://xxx.lanl.gov/abs/hep-th/0105137}{{\tt hep-th/0105137}}].

\bibitem{oai:arXiv.org:hep-th/0111231}
M.~Luty and R.~Sundrum, {\it {Anomaly mediated supersymmetry breaking in
  four-dimensions, naturally}},  {\em Phys.Rev.} {\bf D67} (2003) 045007,
  [\href{http://xxx.lanl.gov/abs/hep-th/0111231}{{\tt hep-th/0111231}}].

\bibitem{Meade:2008wd}
P.~Meade, N.~Seiberg, and D.~Shih, {\it {General Gauge Mediation}},  {\em
  Prog.Theor.Phys.Suppl.} {\bf 177} (2009) 143--158,
  [\href{http://xxx.lanl.gov/abs/0801.3278}{{\tt arXiv:0801.3278}}].

\bibitem{Dimopoulos:1996ig}
S.~Dimopoulos and G.~Giudice, {\it {Multimessenger theories of gauge mediated
  supersymmetry breaking}},  {\em Phys.Lett.} {\bf B393} (1997) 72--78,
  [\href{http://xxx.lanl.gov/abs/hep-ph/9609344}{{\tt hep-ph/9609344}}].

\bibitem{Argurio:2012qt}
R.~Argurio and D.~Redigolo, {\it {Tame D-tadpoles in gauge mediation}},  {\em
  JHEP} {\bf 1301} (2013) 075, [\href{http://xxx.lanl.gov/abs/1206.7037}{{\tt
  arXiv:1206.7037}}].

\bibitem{Komargodski:2009pc}
Z.~Komargodski and N.~Seiberg, {\it {Comments on the Fayet-Iliopoulos Term in
  Field Theory and Supergravity}},  {\em JHEP} {\bf 0906} (2009) 007,
  [\href{http://xxx.lanl.gov/abs/0904.1159}{{\tt arXiv:0904.1159}}].

\bibitem{Komargodski:2010rb}
Z.~Komargodski and N.~Seiberg, {\it {Comments on Supercurrent Multiplets,
  Supersymmetric Field Theories and Supergravity}},  {\em JHEP} {\bf 1007}
  (2010) 017, [\href{http://xxx.lanl.gov/abs/1002.2228}{{\tt
  arXiv:1002.2228}}].

\bibitem{Giudice:1998xp}
G.~F. Giudice, M.~A. Luty, H.~Murayama, and R.~Rattazzi, {\it {Gaugino mass
  without singlets}},  {\em JHEP} {\bf 9812} (1998) 027,
  [\href{http://xxx.lanl.gov/abs/hep-ph/9810442}{{\tt hep-ph/9810442}}].

\bibitem{Randall:1998uk}
L.~Randall and R.~Sundrum, {\it {Out of this world supersymmetry breaking}},
  {\em Nucl.Phys.} {\bf B557} (1999) 79--118,
  [\href{http://xxx.lanl.gov/abs/hep-th/9810155}{{\tt hep-th/9810155}}].

\bibitem{Bagger:1999rd}
J.~A. Bagger, T.~Moroi, and E.~Poppitz, {\it {Anomaly mediation in supergravity
  theories}},  {\em JHEP} {\bf 0004} (2000) 009,
  [\href{http://xxx.lanl.gov/abs/hep-th/9911029}{{\tt hep-th/9911029}}].

\bibitem{Weinberg:2000}
S.~Weinberg, {\it {The quantum theory of fields. Vol. 3: Supersymmetry}},  {\em
  Cambridge University Press} (2000) Cambridge, UK.

\bibitem{Festuccia:2011ws}
G.~Festuccia and N.~Seiberg, {\it {Rigid Supersymmetric Theories in Curved
  Superspace}},  {\em JHEP} {\bf 1106} (2011) 114,
  [\href{http://xxx.lanl.gov/abs/1105.0689}{{\tt arXiv:1105.0689}}].

\bibitem{Dumitrescu:2012at}
T.~T. Dumitrescu and G.~Festuccia, {\it {Exploring Curved Superspace (II)}},
  {\em JHEP} {\bf 1301} (2013) 072,
  [\href{http://xxx.lanl.gov/abs/1209.5408}{{\tt arXiv:1209.5408}}].

\bibitem{D'Eramo:2012qd}
F.~D'Eramo, J.~Thaler, and Z.~Thomas, {\it {The Two Faces of Anomaly
  Mediation}},  {\em JHEP} {\bf 1206} (2012) 151,
  [\href{http://xxx.lanl.gov/abs/1202.1280}{{\tt arXiv:1202.1280}}].

\bibitem{D'Eramo:2013mya}
F.~D'Eramo, J.~Thaler, and Z.~Thomas, {\it {Anomaly Mediation from Unbroken
  Supergravity}},  {\em JHEP} {\bf 1309} (2013) 125,
  [\href{http://xxx.lanl.gov/abs/1307.3251}{{\tt arXiv:1307.3251}}].

\bibitem{Nakayama:2013wda}
Y.~Nakayama, {\it {Consistency of local renormalization group in d=3}},
  \href{http://xxx.lanl.gov/abs/1307.8048}{{\tt arXiv:1307.8048}}.

\bibitem{Baume:2014rla}
F.~Baume, B.~Keren-Zur, R.~Rattazzi, and L.~Vitale, {\it {The local
  Callan-Symanzik equation: structure and applications}},
  \href{http://xxx.lanl.gov/abs/1401.5983}{{\tt arXiv:1401.5983}}.

\bibitem{Dine:2007me}
M.~Dine and N.~Seiberg, {\it {Comments on quantum effects in supergravity
  theories}},  {\em JHEP} {\bf 0703} (2007) 040,
  [\href{http://xxx.lanl.gov/abs/hep-th/0701023}{{\tt hep-th/0701023}}].

\bibitem{Dine:2013nka}
M.~Dine and P.~Draper, {\it {Anomaly Mediation in Local Effective Theories}},
  \href{http://xxx.lanl.gov/abs/1310.2196}{{\tt arXiv:1310.2196}}.

\bibitem{oai:arXiv.org:1104.4482}
Y.~Imamura, {\it {Relation between the 4d superconformal index and the $S^3$
  partition function}},  {\em JHEP} {\bf 1109} (2011) 133,
  [\href{http://xxx.lanl.gov/abs/1104.4482}{{\tt arXiv:1104.4482}}].

\bibitem{oai:arXiv.org:1210.5195}
O.~Aharony, M.~Berkooz, D.~Tong, and S.~Yankielowicz, {\it {Confinement in
  Anti-de Sitter Space}},  {\em JHEP} {\bf 1302} (2013) 076,
  [\href{http://xxx.lanl.gov/abs/1210.5195}{{\tt arXiv:1210.5195}}].

\bibitem{Gripaios:2008rg}
B.~Gripaios, H.~D. Kim, R.~Rattazzi, M.~Redi, and C.~Scrucca, {\it {Gaugino
  mass in AdS space}},  {\em JHEP} {\bf 0902} (2009) 043,
  [\href{http://xxx.lanl.gov/abs/0811.4504}{{\tt arXiv:0811.4504}}].

\bibitem{ArkaniHamed:1997mj}
N.~Arkani-Hamed and H.~Murayama, {\it {Holomorphy, rescaling anomalies and
  exact beta functions in supersymmetric gauge theories}},  {\em JHEP} {\bf
  0006} (2000) 030, [\href{http://xxx.lanl.gov/abs/hep-th/9707133}{{\tt
  hep-th/9707133}}].

\bibitem{Dine:2011gd}
M.~Dine, G.~Festuccia, L.~Pack, C.-S. Park, L.~Ubaldi, et~al., {\it
  {Supersymmetric QCD: Exact Results and Strong Coupling}},  {\em JHEP} {\bf
  1105} (2011) 061, [\href{http://xxx.lanl.gov/abs/1104.0461}{{\tt
  arXiv:1104.0461}}].

\bibitem{Yonekura:2012uk}
K.~Yonekura, {\it {On the Trace Anomaly and the Anomaly Puzzle in N=1 Pure
  Yang-Mills}},  {\em JHEP} {\bf 1203} (2012) 029,
  [\href{http://xxx.lanl.gov/abs/1202.1514}{{\tt arXiv:1202.1514}}].

\bibitem{Novikov:1983uc}
V.~Novikov, M.~A. Shifman, A.~Vainshtein, and V.~I. Zakharov, {\it {Exact
  Gell-Mann-Low Function of Supersymmetric Yang-Mills Theories from Instanton
  Calculus}},  {\em Nucl.Phys.} {\bf B229} (1983) 381.

\bibitem{Shifman:1986zi}
M.~A. Shifman and A.~Vainshtein, {\it {Solution of the Anomaly Puzzle in SUSY
  Gauge Theories and the Wilson Operator Expansion}},  {\em Nucl.Phys.} {\bf
  B277} (1986) 456.

\end{thebibliography}\endgroup

\end{document}